\documentclass{IEEEoj}
\usepackage{cite}
\usepackage{amsmath,amssymb,amsfonts}
\usepackage{algorithmic}
\usepackage{graphicx,color}
\usepackage{textcomp}
\usepackage{tabularx} 
\usepackage{booktabs} 
\usepackage[colorlinks=true,
linkcolor=blue,
citecolor=blue,
urlcolor=blue]{hyperref}

\usepackage{tabularx}
\usepackage{booktabs}
\usepackage{array} 
\def\BibTeX{{\rm B\kern-.05em{\sc i\kern-.025em b}\kern-.08em
    T\kern-.1667em\lower.7ex\hbox{E}\kern-.125emX}}
\AtBeginDocument{\definecolor{ojcolor}{cmyk}{0.93,0.59,0.15,0.02}}
\def\OJlogo{\vspace{-14pt}\includegraphics[height=28pt]{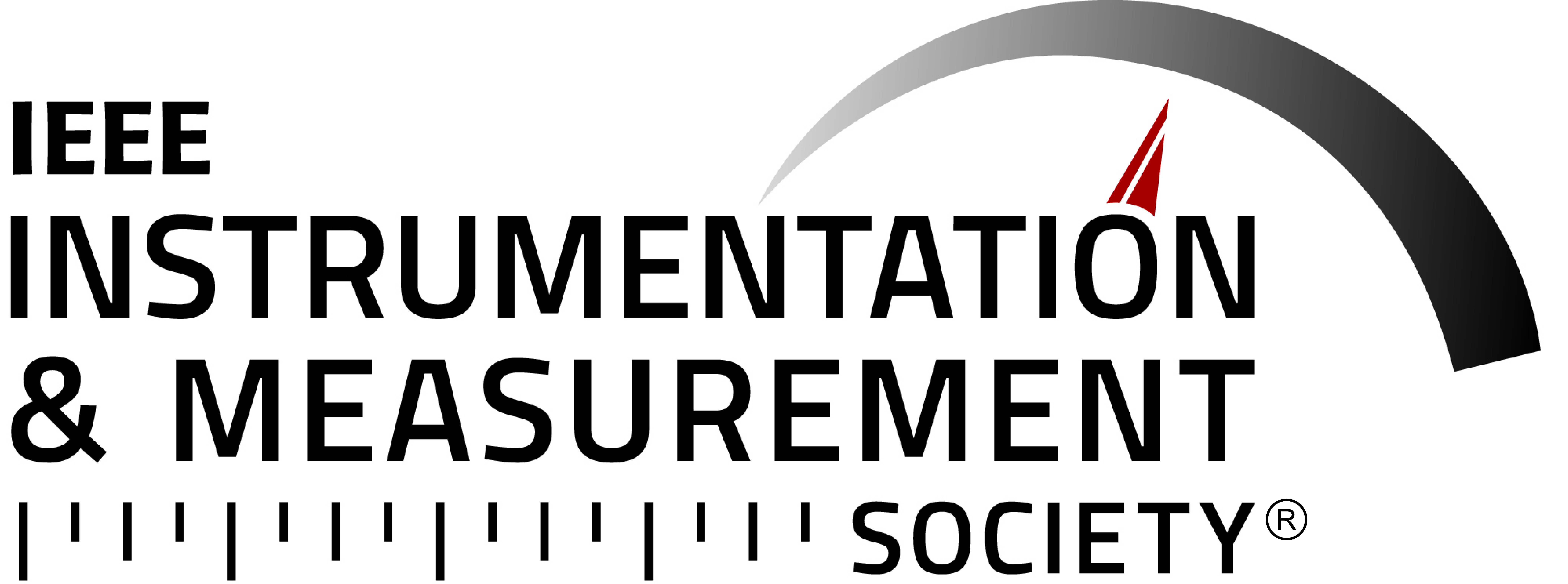}}
\begin{document}
\receiveddate{XX Month, XXXX}
\reviseddate{XX Month, XXXX}
\accepteddate{XX Month, XXXX}
\publisheddate{XX Month, XXXX}
\currentdate{XX Month, XXXX}
\doiinfo{OJIM.2022.1234567}

\title{
	Semblance Distance Analysis for Channel Quality Assessment in Distributed Acoustic Sensing}
\author{
	REINHARDT RADING$^{\textbf{1*}}$, (Member, IEEE),
	MARKUS HOFFMANN$^{\textbf{2}}$,
	NORBERT MEYNERS$^{\textbf{2}}$,
	HOLGER SCHLARB$^{\textbf{2}}$,
	CÉLINE HADZIIOANNOU$^{\textbf{3}}$,
	OLIVER GERBERDING$^{\textbf{4}}$ and
	KATHARINA-SOPHIE ISLEIF$^{\textbf{1,2}}$
}
\affil{Helmut-Schmidt-Universität/Universität der Bundeswehr Hamburg, Holstenhofweg 85, Hamburg, 22043, Germany}
\affil{Deutsches Elektronen-Synchrotron, Notkestraße 85, Hamburg, 22607, Germany}
\affil{University of Hamburg, Institute of Geophysics, Bundesstrasse 55, Hamburg, 20146,Germany}
\affil{University of Hamburg, Institute of Experimental Physics, Luruper Chaussee 149, Hamburg, 22761, Germany}

\corresp{CORRESPONDING AUTHOR: Reinhardt Rading (e-mail: r.r.rading@ieee.org).}

\markboth{Preparation of Papers for IEEE OPEN JOURNALS}{Author \textit{et al.}}

\begin{abstract}
Distributed Acoustic Sensing (DAS) enables dense, kilometer-scale measurements for seismic and environmental monitoring, but its performance is limited by spatially heterogeneous noise and fiber to ground coupling. Identifying high-quality sensing channels remains a challenge, particularly in nonstationary and complex noise environments where conventional signal to noise ratio (SNR) or correlation based approaches are insufficient. In this work, we introduce a statistical framework for DAS channel quality assessment based on a semblance distance metric, which quantifies the deviation of individual channels from a locally coherent reference signal. Unlike traditional pairwise similarity measures, the proposed method formulates channel selection as a distance estimation problem in a coherence enhanced signal space, enabling identification of incoherent noise and coupling induced distortions without requiring prior knowledge of installation conditions or noise statistics. The method is validated using multiple independent datasets, including recordings of the 2023 Turkish-Syrian earthquakes, a controlled seismic source experiment, and long-term continuous measurements. Across all scenarios, the approach identifies regions of low noise and strong coupling, with results that correlate with independent SNR and root mean square error (RMSE) metrics. Furthermore, the method detects subtle sub-decibel temporal variations in coupling conditions over seasonal timescales, demonstrating its capability for continuous fiber health monitoring. These results establish semblance distance analysis as a physically interpretable tool for DAS channel selection with direct applications in seismic imaging, structural health monitoring, and noise mitigation in high sensitivity systems such as gravitational wave detectors.

\end{abstract}



\maketitle

\section{INTRODUCTION}
\IEEEPARstart{D}{istributed} Acoustic Sensing (DAS) is an emerging fiber optic sensing technology that enables standard telecommunication optical fibers to function as dense, kilometer-scale arrays for measuring seismic and environmental perturbations. DAS measures axial strain or strain rate along an optical fiber by detecting phase changes in Rayleigh-backscattered light induced by external perturbations \cite{b9}. Variations in optical phase arise from changes in fiber length and refractive index, which are caused by physical processes such as dynamic strain, temperature fluctuations, pressure variations, and mechanical stress\cite{b10}.

In recent years, DAS has been applied across a broad range of domains, including infrastructure monitoring in the oil and gas and telecommunications industries\cite{t0,t1,t2}, seismic imaging\cite{t3}, and environmental monitoring in seismology\cite{a1}. Building on these advances, DAS is now opening new frontiers in applications such as linear infrastructure monitoring\cite{a0}\cite{a00}, noise mitigation in precision research facilities\cite{b1}, large scale environmental monitoring\cite{b1}, and prospective deployment in future lunar missions\cite{b2}. These opportunities are driven by several  advantages of DAS, including low deployment cost and capability to operate in harsh or inaccessible environments. Another advantage of the DAS system over conventional seismic sensors is that it transforms the entire fiber length into several virtual sensor points. The number of sensors is defined by the system’s spatial resolution. Therefore, one can choose sensors with the best  signal to noise ratio (SNR) and strong coupling, while abandoning sensors contaminated with noise and poor coupling. Spatial resolution is the minimum distance over which two independent measurement values can be distinguished. It is determined by the interrogating laser pulse width\cite{b10}.

Despite its numerous advantages, DAS faces challenges that limit its performance in emerging high sensitivity applications. In addition to measurand induced phase changes, the measured DAS signal is influenced by several noise sources including fiber length fluctuations associated with linear propagation effects, nonlinear Kerr-induced phase noise\cite{b10}\cite{b6}, and coupling effects\cite{d1,d2,d3}, and other instrumental effects. Such noise vary spatially along the fiber. Furthermore, due to its broadband response (between 15 mHz and 450 Hz), DAS records a wide range of signals within its operational bandwidth, including both anthropogenic and environmental noise. This is particularly evident in urban and industrial environments, where noise sources are often directional, spatially heterogeneous, and time varying, with nonstationary spectral characteristics. Such complexity makes it difficult to distinguish between signal and noise using conventional signal processing techniques. In particular, these noise sources can exhibit both local and global behavior, possess unknown characteristics, and lack well defined spectral signatures which complicates their identification and mitigation; an essential requirement for high sensitivity applications. For instance, in applications such as noise mitigation in research facilities \cite{b1}\cite{b4} and urban seismic imaging \cite{b5}, the selection of channels with high  SNR and minimal noise contamination is very important.

\begin{figure}
 	\centerline{\includegraphics[width=3.5in]{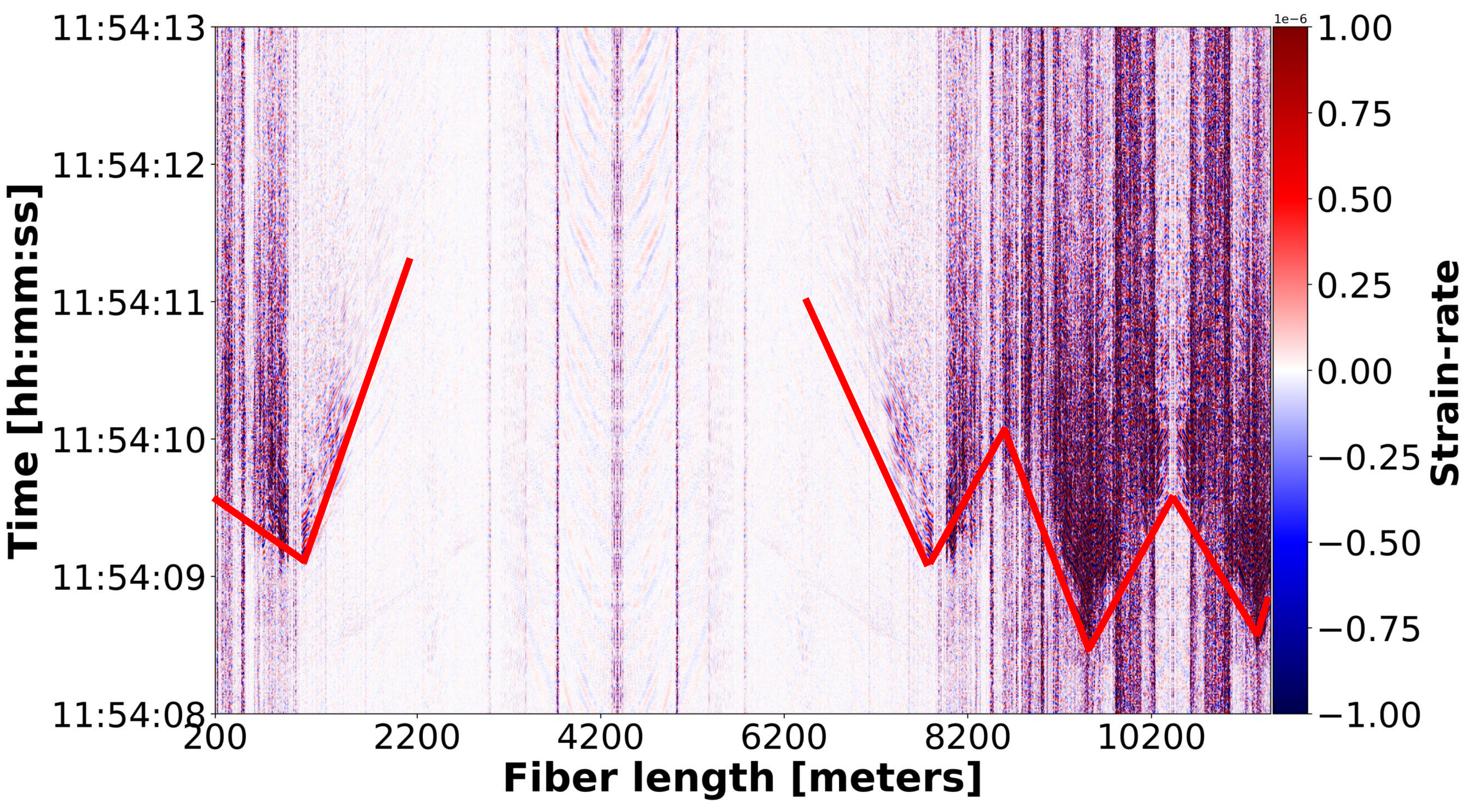}}
 	\caption{Waterfall plot illustrating how different sections of the fiber respond to the same experiment. Certain segments exhibit stronger coupling and preserve signal integrity, clearly recording the event, while others show little to no response. This variation highlights how coupling and local fiber conditions influence whether specific channels can effectively capture the seismic signal.\label{fig3}}
 \end{figure}

\begin{figure}
 	\centerline{\includegraphics[width=3.5in]{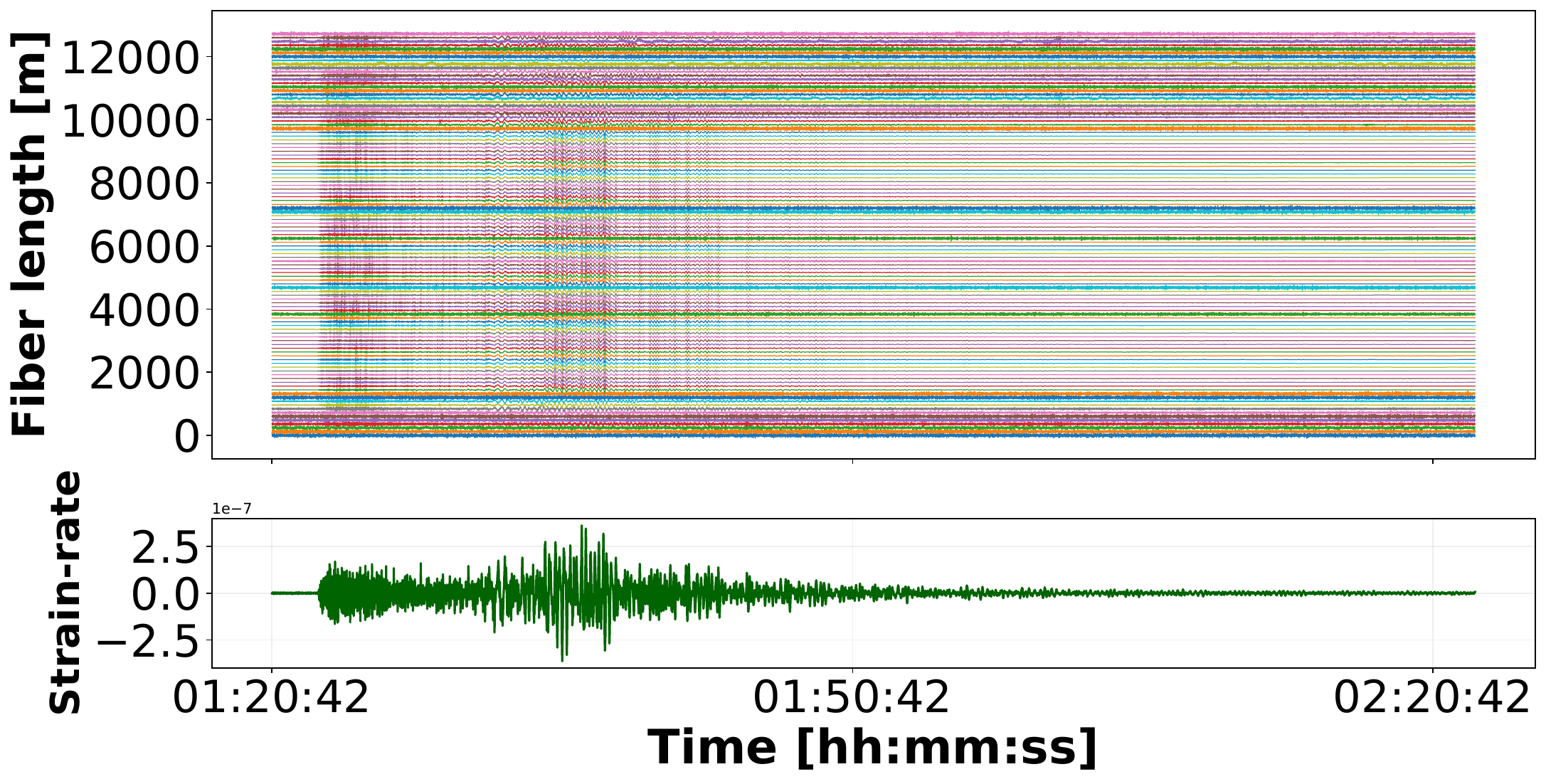}}
 	\caption{The 12.5 km dark fiber network in Hamburg captured signals from the 7.8-magnitude Turkish-Syrian earthquake on February 6, 2023. The plot illustrates the response of different fiber segments to the event. Clear earthquake signals are observed along the 1,200 to 8,500 m section, whereas segments below 1,200 m and beyond 10,000 m are noise dominated and do not effectively record the event.\label{fig90}}
 \end{figure}

An additional challenge arises from the variability in coupling between the optical fiber and the surrounding medium \cite{d1}\cite{d2}. In most deployments, optical fibers are buried underground under conditions that are spatially heterogeneous and often unknown, making direct inspection and calibration impractical. As a result, coupling conditions vary along the fiber length, leading to channel to channel differences in sensitivity\cite{d3}. Well coupled sections of the fiber are more likely to accurately capture seismic or environmental strain, whereas poorly coupled segments show reduced sensitivity and distorted responses \cite{d1,d2,d3} as shown in Figure \ref{fig3} and \ref{fig90}. These discrepancies are further worsened by the presence of multiple protective coating layers surrounding the fiber, which mechanically isolate the glass core from the host material. While these coatings enhance durability, they also introduce deviations between the true ground strain and the strain experienced by the fiber core. The strain transfer through these layers is nonlinear and frequency dependent, resulting in variations in both amplitude and phase response across channels \cite{b3}. 

Improving DAS performance in high sensitivity applications therefore requires excellent methods to identify channels with high SNR while neglecting those degraded by poor coupling or noise contamination. This is important in applications such as seismic wavefield reconstruction, where data from high-SNR selected channels can be used to minimize noise in the reconstructed wavefield. Another application is noise cancellation for gravitational wave detectors, where performance depends directly on the quality of the recorded data\cite{b4}. It is also relevant to future planetary and forensic seismology, where weak signals must be reliably detected and interpreted.

 A variety of methods have been proposed to identify low-quality channels, typically based on cross-correlation \cite{e10}\cite{e11}, coherence analysis\cite{e20}\cite{e21}, or channel dependent SNR estimation\cite{e30}\cite{e31}. While effective under controlled conditions, these approaches may be insufficient for applications requiring high sensitivity. In particular, they may lack sensitivity to detect very minute variations, fail to capture nonlinear noise effects, and introduce uncertainties when applied in non-stationary noise environments as will be shown in this study. Conventional SNR estimation methods often rely on assumptions regarding the statistical properties of noise, assuming stationarity \cite{b66,b7,b8}. This assumption is invalid for DAS measurements, where noise can exhibit nonlinear and time varying behavior\cite{b6}, thereby limiting the reliability of such methods in high-precision applications.
 
 In this study, we introduce a statistical framework based on a semblance distance metric that extends beyond conventional coherence and SNR based methodologies by quantifying inter-channel dissimilarity within DAS arrays. In contrast to existing techniques, the proposed approach captures non-linear distortion and non-stationary noise behavior, enabling the identification of high SNR channels, as well as the detection of fiber segments exhibiting poor ground coupling. 
 The distinguishing feature of this method is the formulation of channel quality assessment as a distance estimation problem through the use of a semblance distance metric. Rather than relying on pairwise similarity measures, the approach constructs a locally coherent reference signal and evaluates channel-wise deviations from this reference. This enables the detection of incoherent noise contributions arising from variable coupling conditions. The method relies on the statistical properties of recorded signals, therefore it remains unbiased to the signal type, excitation source, and energy scale. Table I summarizes the differences between existing methodologies and the semblance-distance framework introduced in this work.

 \begin{table}[t]
 	\caption{Comparison of Methods}
 	\label{tab1}
 	\centering
 	\footnotesize
 	\renewcommand{\arraystretch}{1.1}
 	\setlength{\tabcolsep}{3pt}
 	
 	\begin{tabularx}{\columnwidth}{|>{\raggedright\arraybackslash}X
 			|>{\raggedright\arraybackslash}X
 			|>{\raggedright\arraybackslash}X
 			|>{\raggedright\arraybackslash}X
 			|>{\raggedright\arraybackslash}X|}
 		\hline
 		\textbf{Method} & \textbf{Idea} & \textbf{Strengths} & \textbf{Limitations} & \textbf{Semblance distance contribution} \\
 		\hline
 		
 		SNR-based \cite{e30}\cite{e31} & Estimate signal quality:
 		signal power and noise power & Simple & Needs noise estimate; assumes stationarity; window sensitive & \textbf{No noise model; good for nonstationary data} \\
 		\hline
 		
 		Cross-correlation\cite{e10}\cite{e11} or coherence \cite{e20}\cite{e21} & Channel similarity & Captures spatial coherence & Pairwise only; sensitive to phase shifts \& amplitude variations; scales poorly for large arrays & \textbf{Uses a global coherent reference instead of pairwise comparison} \\
 		\hline
 		
 		Clustering or PCA\cite{kk1}\cite{kk2} & Anomaly detection & Captures global structure & Low interpretability; tuning required & \textbf{Physically interpretable metric tied to signal coherence and coupling behavior} \\
 		\hline
 		
 		Thresholding based filtering\cite{kk3} & Rule-based filtering & Simple implementation & Manual tuning; non-adaptive & \textbf{Automatic deviation detection} \\
 		\hline
 		
 		\textbf{Semblance Distance (Proposed method)} & \textbf{Distance from a coherent reference in a coherence signal space} & \textbf{Scalable; no noise assumptions; robust to diverse environments} & \textbf{Requires selection of spatial window and reference estimation} & \textbf{--} \\
 		\hline
 	\end{tabularx}
 	\label{table1}
 \end{table}

 The remainder of this paper is structured as follows: Section II introduces the theoretical framework describing the proposed methodology; Section III details the approach for identifying low-noise and poorly coupled channels; Section IV describes the experimental setup, including the datasets and events used for validation; Sections V and VI present and discuss the experimental results and the application of the metric; finally, Section VII summarizes the main findings and outlines directions for future research.

\section{THEORY}
For a fiber of total length $L$ and spatial resolution $k$, the number of DAS sensors (interchangeably used here as channels) is \begin{math}
	N=\frac{L}{k}.
\end{math} Consider an array of $N$ sensing channels, distributed along an optical fiber of length $L$, with sensor locations denoted as $\{S_{1},S_{2},S_{3}\cdots S_{N}\}$ where $S_{1}$ corresponds to the first channel and $S_{N}$ to the last. For a transient event (e.g., an earthquake), the DAS system records a data matrix:  $\mathbf{X} \in \mathbb{R}^{N \times M}$ where $M$ denotes the number of temporal samples per channel. Each channel $i$ records a signal and the noise described by:
\begin{equation}
	Y_{i}(t)= X_{i}(t)+N_{i}(t), 
	\label{eq1}
\end{equation} where \begin{math}
N_{i} (t)=N_{c}(t)+N_{in}(t).
\end{math} The noise term $N_{i}$ consists of two contributions: local coherent noise $N_{c}$ and channel-dependent incoherent noise $N_{in}$, The channel dependent output $Y_{i}$
is a combination of:
\begin{enumerate}
	\item the coherent event related signal, $X_{i}$,
	\item spatially varying unwanted coherent local noise, $N_{c}$,
	\item  channel dependent incoherent instrumental and coupling noise, $N_{in}$.
\end{enumerate}

The unwanted local coherent noise may originate from localized disturbances near portions of the fiber and therefore does not appear coherently across the array and can be estimated and suppressed using the modified semblance method. By extending classical semblance \cite{c1,c2,c3,c4}, we developed a modified semblance operator that incorporates Hilbert-transform phase information for coherent reference construction given by:
	\begin{equation}
	A_{ref}(t)=\frac{\sum_{i=1}^{N} |f_{i}(t)|^2 + \sum_{i=1}^{N} |h_{i}(t)|^2}{N \sum_{i=1}^{N} \{f_{i}(t)^2 + h_{i}(t)^2\}},
	\label{eq2}
\end{equation}

where $A_{ref}(t)$ is a temporal sample of the locally coherent reference signal, $f_{i}(t)$ is the timeseries of sensor $i$ and $h_{i}(t)$ is the Hilbert transform of $f_{i}(t)$. 

The method enables suppression of spatially unwanted coherent environmental noise while preserving phase distortions caused by poor coupling or instrumental noise; the distinguishing feature of our approach. It also operates as a spatial filter that enhances signal components that are consistent across the array while suppressing incoherent contributions. The output of (\ref{eq2}) yields a time series representing a local noise reduced estimate of the coherent signal. This reference signal can then be used to correct individual channel recordings through scaling or normalization relative to the semblance derived output. To preserve signal content at a given seismic wavelength, the channels selected must be comparable to or greater than the wavelength of interest to avoid losing relevant signal information\cite{c4}.


After suppression of spatially varying unwanted environmental noise, the remaining disturbances consist of instrumental noise and channel-dependent coupling effects. These noise are difficult to remove through spatial coherence filtering because they are uncorrelated and vary unpredictably between channels. The channels with these remaining disturbances can be identified using the \textbf{Semblance distance} (\textbf{SemD}) method. \textit{Distance} here does not refer to physical separation along the fiber, but rather to the dissimilarity between two time series. A distance value of zero indicates that two time series are identical (highly similar), while larger values indicate increasing dissimilarity due to noise or coupling artifacts.

The SemD between two time-series is given as:
\begin{equation}
	d(A_{ref},B_{i})=\sqrt{\sum_{t=1}^{M} \left| A_{\mathrm{ref}}(t)-B_{i}(t) \right|^2},
	\label{eq3}
\end{equation} where 
\begin{equation}
B_{i}(t)=\frac{N \{|f_{i}(t)|^2 + |h_{i}(t)|^2\}}{ \sum_{i=1}^{N} \{f_{i}(t)^2 + h_{i}(t)^2\}},
	\label{eq4}
\end{equation} is the semblanced-based time series in a DAS array.

Alternative distance metrics, such as cosine \cite{c01} or correlation distance\cite{c02}, are invariant to amplitude scaling, hence less sensitive to coupling-induced attenuation, which is the phenomenon of interest in DAS and this paper. Mahalanobis distance\cite{c03} requires the estimation of a large covariance matrix, which becomes computationally prohibitive for arrays containing thousands of channels. In contrast, Wasserstein distance\cite{c04} compares probability distributions rather than time-domain waveform agreement. By comparison, Euclidean distance\cite{c044} is a computationally efficient and physically interpretable measure of sample-by-sample deviation from the coherent reference signal. This makes it well suited to identifying channels affected by poor coupling or elevated noise.

\vspace{0.25cm}
\vspace{0.45cm}

\vspace{0.25cm}

\noindent{\large \textbf{\textit{How SemD Detects Poor Coupling}}}

\noindent The channel-dependent incoherent noise $N_{in}(t)$ in (\ref{eq1}) comprises of two components: instrumental noise; which can be estimated and subtracted using \cite{c05} and coupling-induced distortion, a non-linear distortion effect arising from strain transfer via ground contact stiffness (modelled as a sinc function\cite{d1,d2,d3}), thermally modulated over time, and spatially heterogeneous due to variable burial depth and installation geometry. 

\noindent The modified semblance operator suppresses the spatially coherent unwanted noise $N_{c}(t)$ by enforcing phase and amplitude alignment across channels, yielding a noise-reduced reference signal $A_{ref}(t)$. After removing the estimated instrumental noise component, the residual channel-dependent deviations from $A_{ref}(t)$ are dominated by coupling-induced effects. Channels with poor coupling produce large SemD values, while well-coupled channels remain phase-aligned and amplitude-coherent with the reference, minimizing SemD values. Coupling effects are nonlinear and spatially heterogeneous, making it impossible to model a priori. It must instead be detected through observed deviations from the coherent signal structure. This explains why conventional SNR-based methods; which assume stationarity and additive noise, fail to detect slow coupling variations, while SemD, which makes no assumptions about the noise model, identifies them.

\vspace{0.3cm}

Consider a DAS data matrix: $N \times M$, to quantify the mutual similarity between sensors, a pairwise semblance based distance measure is computed using (\ref{eq3}). The total number of unique sensor pairs is \begin{math}
	\frac{N(N-1)}{2},
\end{math} which scales with the number of sensors but remains computationally tractable for typical array sizes on standard computing platforms. The accuracy of the SemD estimation depends on the number of temporal samples $M$. Similar to the law of large numbers, increasing $M$ improves statistical reliability and reduces variance in the dissimilarity estimate. However, this improvement comes at a computational cost. If the SemD is computed using a time alignment procedure \cite{s0} (e.g., dynamic time alignment), the temporal computational complexity scales as \begin{math}
O(M^{2}).
\end{math} Thus, increasing the number of time samples  increases computational expense, particularly for large sensor arrays. 

\begin{figure}
	\centerline{\includegraphics[width=3.5in]{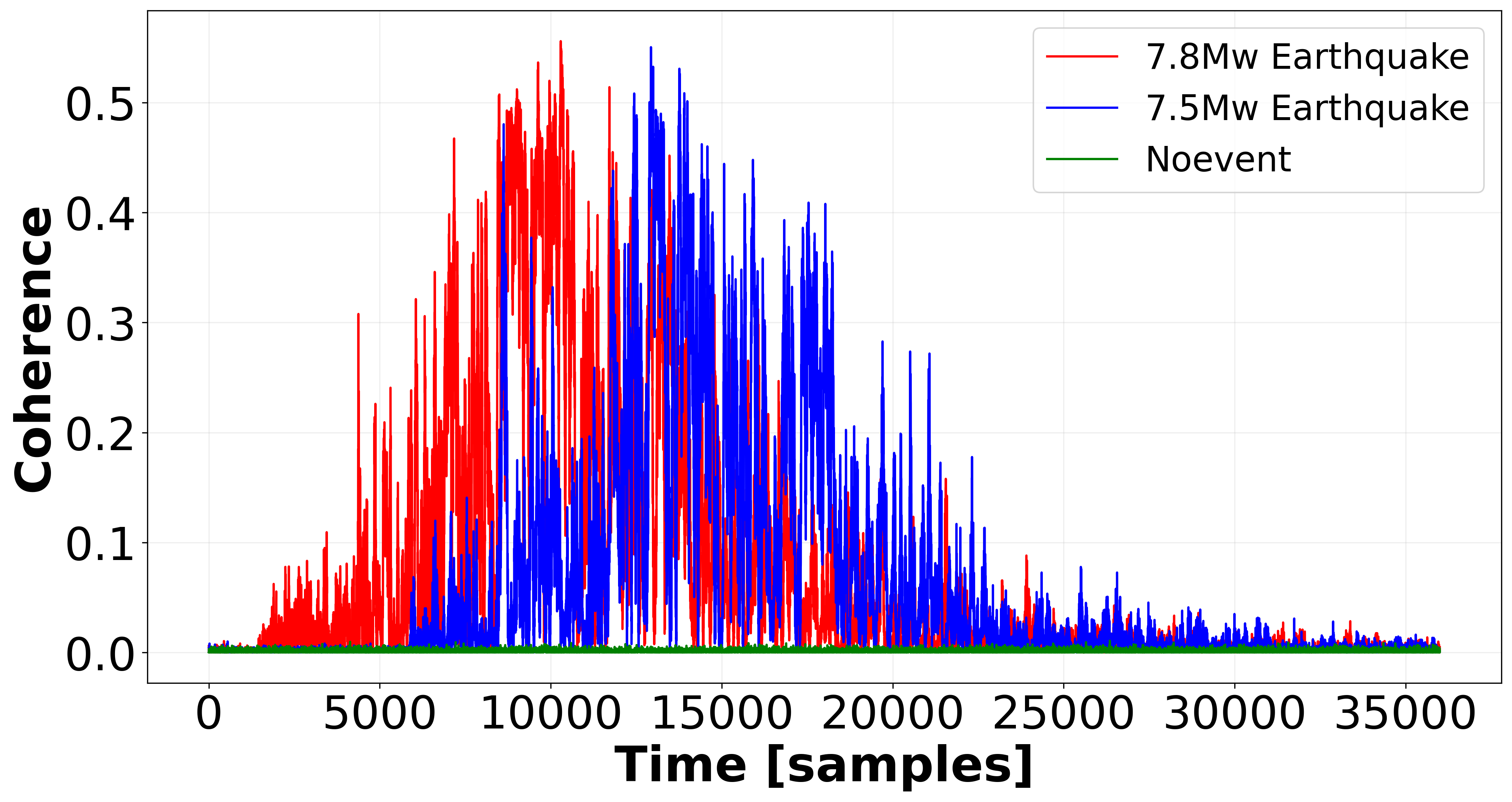}}
	\caption{Timeseries showing local noise-reduced estimates of the coherent signal, quantified using semblance coherence as defined in (\ref{eq2}). The two time series with magnitude above 0.2 correspond to the 7.8Mw and 7.5Mw Turkish-Syria earthquakes that occurred on 6 February 2023. The Noevent time series with magnitude below 0.005 represents a reference dataset recorded at 01:00 UTC on 7 February 2023, the day after the earthquakes, during a period with no significant seismic or anthropogenic activity.\label{fig18}}
\end{figure}

To mitigate the computational cost, constrained alignment strategies should be employed, for example, the use of an Itakura parallelogram \cite{s1} of width \begin{math}
G
\end{math} or a Sakoe-Chiba band\cite{s2} of width \begin{math}
G
\end{math}. As an illustrative case, a four-month DAS dataset with 15000 channels was analyzed using a Sakoe-Chiba band with a width of 4 hours on a high-performance computing cluster. The computation was executed on a single node equipped with 120 CPU cores, with parallelization implemented via OpenMP using 120 threads. To prevent oversubscription, numerical libraries (MKL, OpenBLAS, and NumExpr) were restricted to single-thread execution. Under this configuration, a single simulation required approximately 96 hours to complete. For short-duration applications such as earthquake detection or transient event characterization, where $M$ typically corresponds to seconds to minutes of data, the semblance distance computation remains fast and tractable even for arrays comprising thousands of channels. The 96-hour case represents the opposite extreme: continuous monitoring over a four-month period, where the goal is detecting slow seasonal coupling drift rather than identifying channels for a specific transient event.

The width \begin{math}
G
\end{math} must be selected carefully and justified based on the temporal characteristics of the noise process under investigation. In particular,  \begin{math}
G
\end{math} should neither exceed nor fall below the characteristic time scale of the noise of interest. Choosing \begin{math}
G \ll M
\end{math} ensures a favorable trade-off between computational efficiency and alignment flexibility.

\section{IDENTIFYING LOW NOISE AND POOR COUPLED CHANNELS}

\subsection{Fiber Distance Selection}
A subset of DAS channels is selected within a spatially confined region of interest. For low-frequency events, the effective sensing length should be considered relative to the wavelength of interest. Channels distributed over distances significantly larger than the spatial coherence length of the signal should be avoided, as they may introduce irrelevant contributions. Including such channels only increases computational complexity.

\subsection{Local Coherent Signal Estimation}
After selecting the $N$ channels, incoherent contributions within the chosen fiber segment are suppressed using (\ref{eq2}). This reference signal serves multiple purposes:
\begin{enumerate}
	\item It provides a coherent benchmark against which individual channels can be compared and corrected.
	\item It enables quantitative validation of the algorithm through error metrics such as the root mean square error (RMSE) between each channel and the coherent estimate.
	\item It facilitates the identification of channels whose response deviates from the coherent signal.
	
\end{enumerate}

\subsection{Semblance Distance (SemD) Analysis}
The SemD can be  computed using three approaches.
In the first approach, the semblance distance is calculated directly using (\ref{eq3}), where $A_{ref}(t)$ represents the coherent reference signal obtained from (\ref{eq2}) and $B_{i}(j)$ denotes the semblance transformed time series of the $i$-th DAS channel. Each channel is compared to the locally noise reduced estimate of the coherent signal $A_{ref}$. The resulting metric quantifies the deviation of individual channels from the expected coherent response and produces a one dimensional vector of length 
$N$, characterizing the semblance distance of each channel relative to the reference signal.

In the second approach, all time series of an array is converted using (\ref{eq4}) and then semblance distance computed using a time alignment procedure based on dynamic time alignment\cite{s0}. In this case, all selected channels are compared pairwise after compensating for potential time shifts or small phase misalignments. This procedure yields an $N \times N$ matrix that describes the inter-channel dissimilarity structure within the selected fiber segment. To obtain a channel wise distance from the pairwise analysis, the 
 $N \times N$ matrix is averaged along one axis, resulting in a one dimensional vector of length $N$. This vector represents the relative deviation of each channel from the ensemble behavior of the selected region.

 In the last approach, semblance distance directly uses a time-alignment procedure. This is beneficial in the low-frequency domain, especially for understanding long-time changes in coupling resulting from climatic or infrastructure changes. It slightly outperforms the second approach. Similar to the second other approach, it produces an $N \times N$ matrix, which is then reduced to a one-dimensional vector of length $N$.
 
 \begin{figure}
 	\centerline{\includegraphics[width=3.5in]{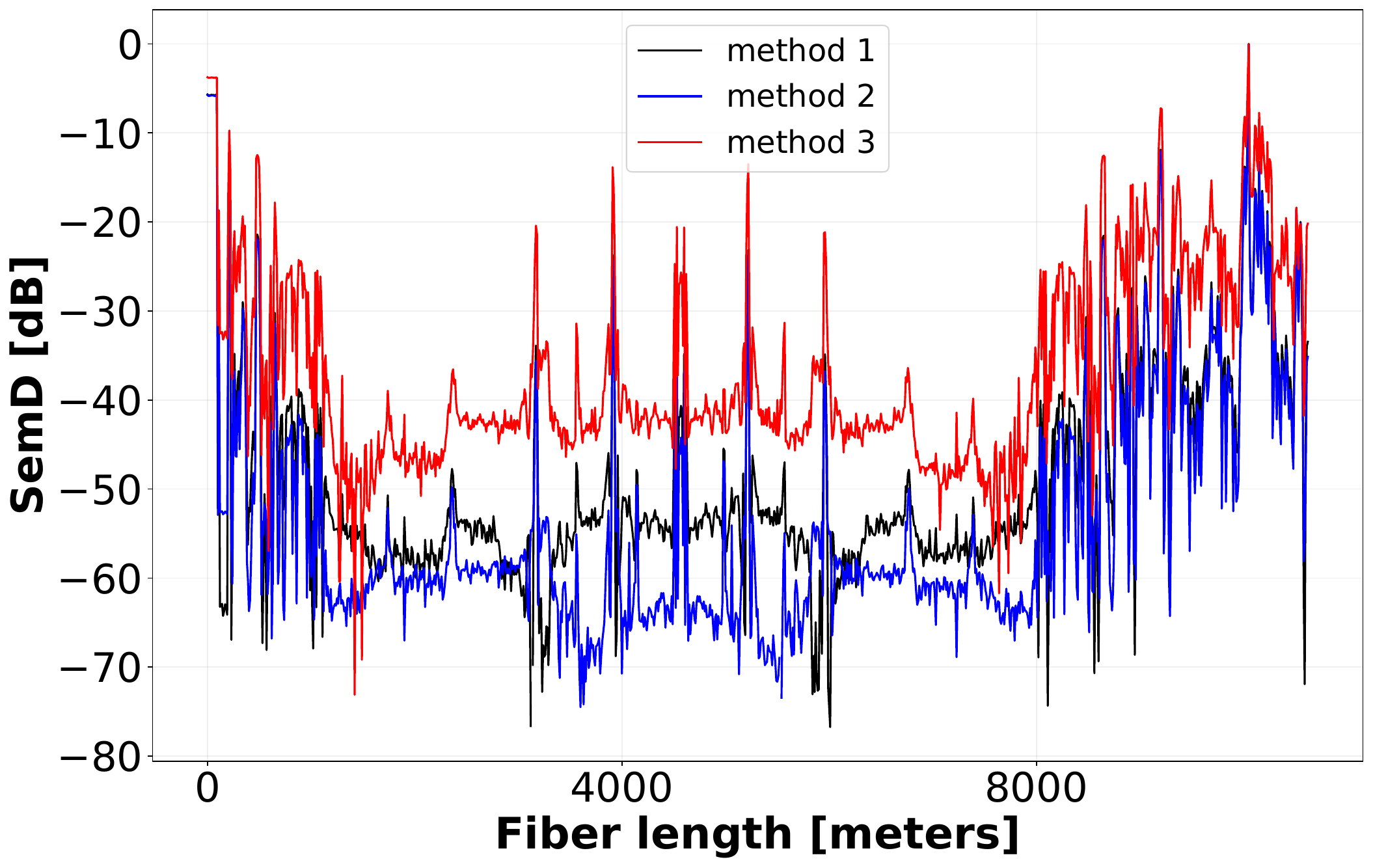}}
 	\caption{The three Semblance Distance (SemD) approaches used to calculate the SemD during the noevent period. The performance of all three approaches is similar, with each successfully identifying channels with both low and high noise levels. The main difference between the approaches is in the magnitude of the calculated SemD, although these differences are relatively small.\label{fig100}}
 \end{figure}

 Channels with larger SemD values exhibit reduced coherence with the coherent reference, indicating elevated noise levels caused by instrumental noise (if not subtracted) or poor mechanical coupling to the surrounding medium. Conversely, channels with values approaching zero demonstrate strong agreement with the ensemble response and are therefore classified as well coupled channels.

\section{EXPERIMENTAL SETUP}
\begin{figure}
	\centerline{\includegraphics[width=3.5in]{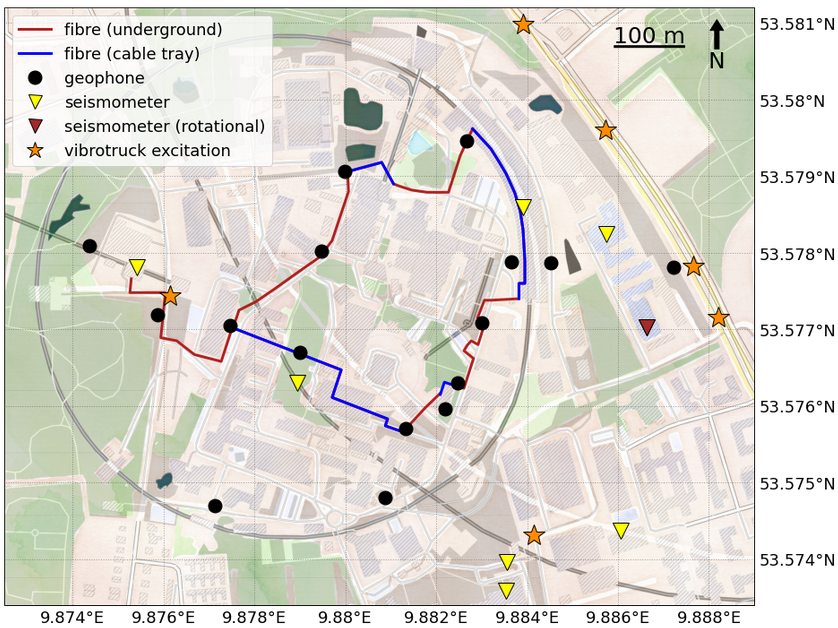}}
	\caption{Map of a segment of the 12.5 km dark fiber within the WAVE network, highlighting its diverse installation environments. Portions of the fiber run along cable trays, others are buried underground, and some sections are exposed to anthropogenic noise, such as nearby road traffic. This heterogeneity makes the network suited for studying semblance distance metric.\label{fig10}}
\end{figure}

The experimental validation was conducted using data from the WAVE network \cite{b1}. In the network exists a 15 km dark-fiber infrastructure characterized by heterogeneous conditions and diverse coupling environments. The fiber traverses areas with varying levels of anthropogenic and environmental noise, providing an ideal testbed for evaluating the statistical method of identifying low-noise and poorly coupled fiber channels.
The fiber is deployed in multiple configurations: some sections are installed on cable trays inside and outside buildings. Some segments routed through the research campus area exposed to  anthropogenic noise sources e.g., vehicular traffic as shown in Fig. \ref{fig10}. Some fiber sections are located in tunnels with some section of the tunnel buried approximately 20 m underground, where the surrounding soil acts as a mechanical low-pass filter. The differing environments result in spatially varying coupling conditions, including sections with partially unknown or variable ground coupling. 

The fiber is interrogated using a DAS interrogator with a spatial resolution of 1 m, yielding more than 15,000 virtual sensing channels along the fiber. The system continuously records a broad range of signals\cite{b1}, including natural events (e.g., earthquakes and thunderstorms) as well as anthropogenic disturbances.

To evaluate the proposed algorithm, we selected events that were coherently recorded along the full length of the fiber. These included the twin Turkish-Syrian earthquakes of 6 February 2023 and a controlled stone-drop experiment. For the stone-drop event, the DAS data were initially acquired at a sampling frequency of 1000 Hz and subsequently downsampled by a factor of 10 to 100 Hz. A 5-minute time segment was used in the analysis. For the earthquake events, the data were further downsampled to 10 Hz, and a 60 minute time window was analyzed to assess the algorithm’s performance under long duration, low frequency excitation. 

To investigate the temporal stability of the identified low-noise and poorly coupled channels, continuous DAS recordings acquired between February and May 2025 were additionally analyzed with a Sakoe-Chiba width size of 4 hours. The objective was to determine whether fiber ground coupling conditions or fiber noise characteristics show temporal variability over extended periods. For this four month dataset, the data were further downsampled to one sample per minute prior to evaluation. 

To identify low-noise and poorly coupled channels, we followed the procedure described in Section III and validated the results using both SNR and RMSE metrics. The SNR was computed using the variance of the coherent reference signal obtained from (\ref{eq2}) and the variance of the residual obtained by subtracting each individual channel time series from this reference signal. The RMSE was calculated by first computing the mean squared difference between each channel time series and the coherent reference signal, followed by taking the square root of this mean value.

The study focused on the frequency range from sub-Hz frequencies up to 48 Hz. This band encompasses the dominant contributions from environmental excitations and anthropogenic disturbances and is relevant for coupling assessment and noise characterization in DAS systems. 

It is important to note that the fiber network configuration changed during the study period. In 2023, the dark fiber had a total length of 12.5 km. In 2024, the network was extended by 2.5 km, resulting in a total length of 15 km. Consequently, the analyzed fiber distance differs between datasets acquired in 2023 and those recorded in 2025. This variation is explicitly reflected in the Results section, where channel statistics are presented as a function of fiber distance.

\section{RESULTS}

\subsection{TWIN TURKISH-SYRIAN EARTHQUAKES  2023}
We first applied the proposed algorithm to DAS data collected during the twin Turkish-Syrian earthquakes. The first earthquake of magnitude 7.8 occurring at 01:17 UTC, followed by an earthquake of magnitude 7.5 occurring at 10:24 UTC. In addition, we analyzed a reference dataset recorded at 01:00 UTC on February 7, 2023, the day after the earthquakes when no significant seismic or anthropogenic activity was observed. At the time of DAS acquisition, the dark fiber had a total length of 12.5 km.

\begin{figure}
	\centerline{\includegraphics[width=3.5in]{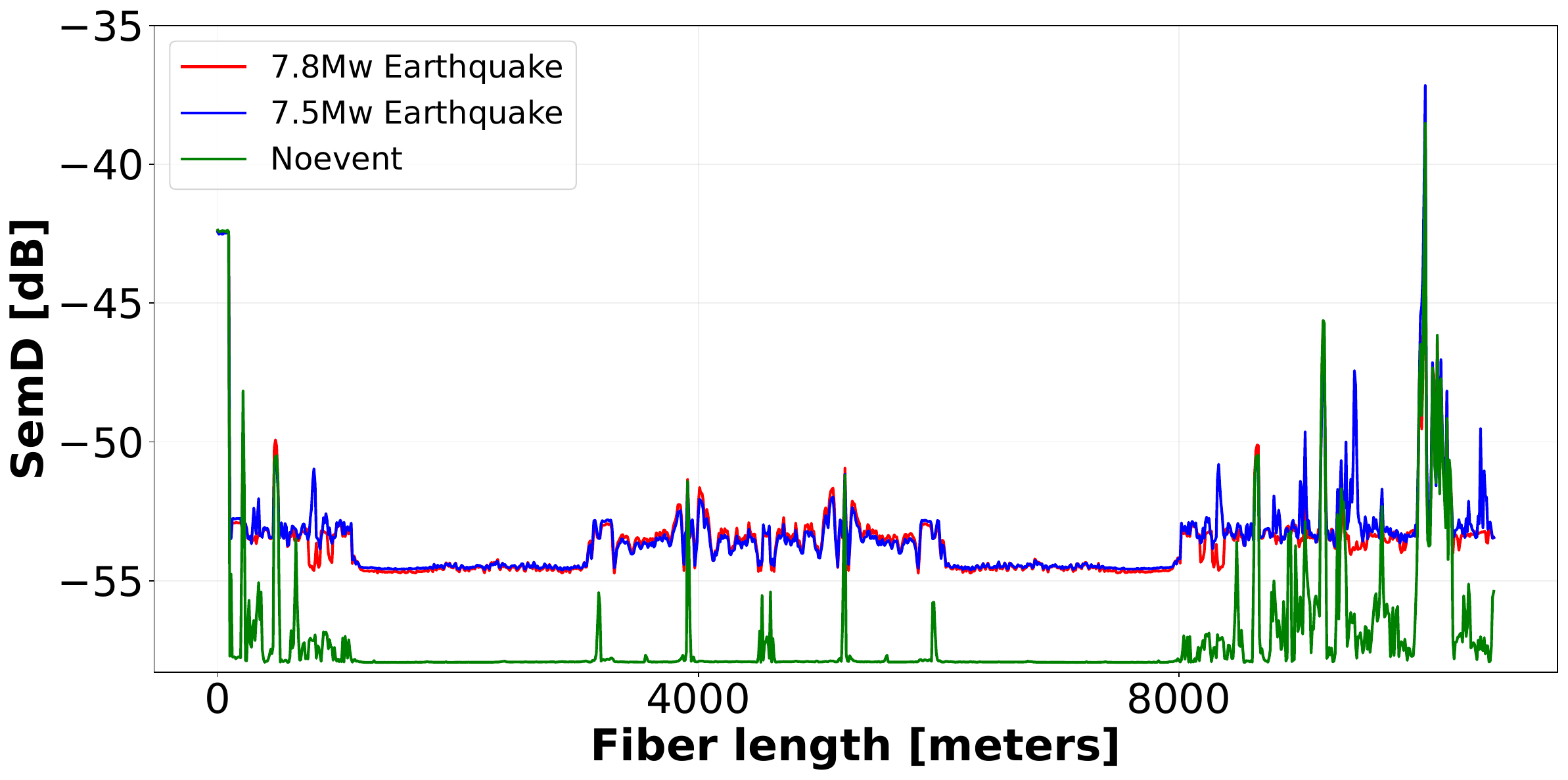}}
	\caption{Semblance distance computed for three cases (7.8 Mw and 7.5 Mw earthquakes, and a no-event scenario). The algorithm identifies the quietest channels along the fiber segment between approximately 1100 m and 8000 m, with an average semblance distance of about -54.6 dB, and detects transitions in coupling conditions. In contrast, the fiber sections from 0 to 1100 m and 8000 to 12,500 m exhibit higher semblance distance values (-53 to -36 dB), indicating elevated noise levels and/or poorer coupling.\label{fig5}}
\end{figure}

The algorithm identifies the quietest channels within the fiber segment extending from approximately 1110 m to 8000 m, with an average semblance distance of about -54.6 dB during the earthquake recordings as shown in Fig. \ref{fig5}. This result is consistent with the known physical deployment of the network: this section of the fiber is buried approximately 20 m underground, where the surrounding soil acts as a mechanical low-pass filter and shields the fiber from surface related anthropogenic disturbances. Within this low-noise region, localized spikes in semblance distance are observed, for example at 3200 m and 4400 m. These correspond to short fiber sections that are not buried and are instead routed in a different environment, where they are exposed to environmental perturbations. The algorithm is able to detect these transitions in coupling conditions.
In contrast, the fiber sections between 0 and 1100 m and between 8000 and 12,500 m show  higher semblance distance values (between -53 and -36 dB), indicating higher  noise levels and or poor coupling. These sections are located within the campus region, where the fiber is installed on cable trays rather than buried underground. Consequently, they are exposed to multiple anthropogenic noise sources, including tree induced vibrations, transformer hum, human-induced noises and vehicular traffic. The high semblance distance in these regions shows both stronger environmental excitation and probable spatially varying mechanical coupling. The algorithm is able to discriminate between low-noise underground segments and poorly coupled, high-noise exposed surface installations.

\begin{figure}
	\centerline{\includegraphics[width=3.5in]{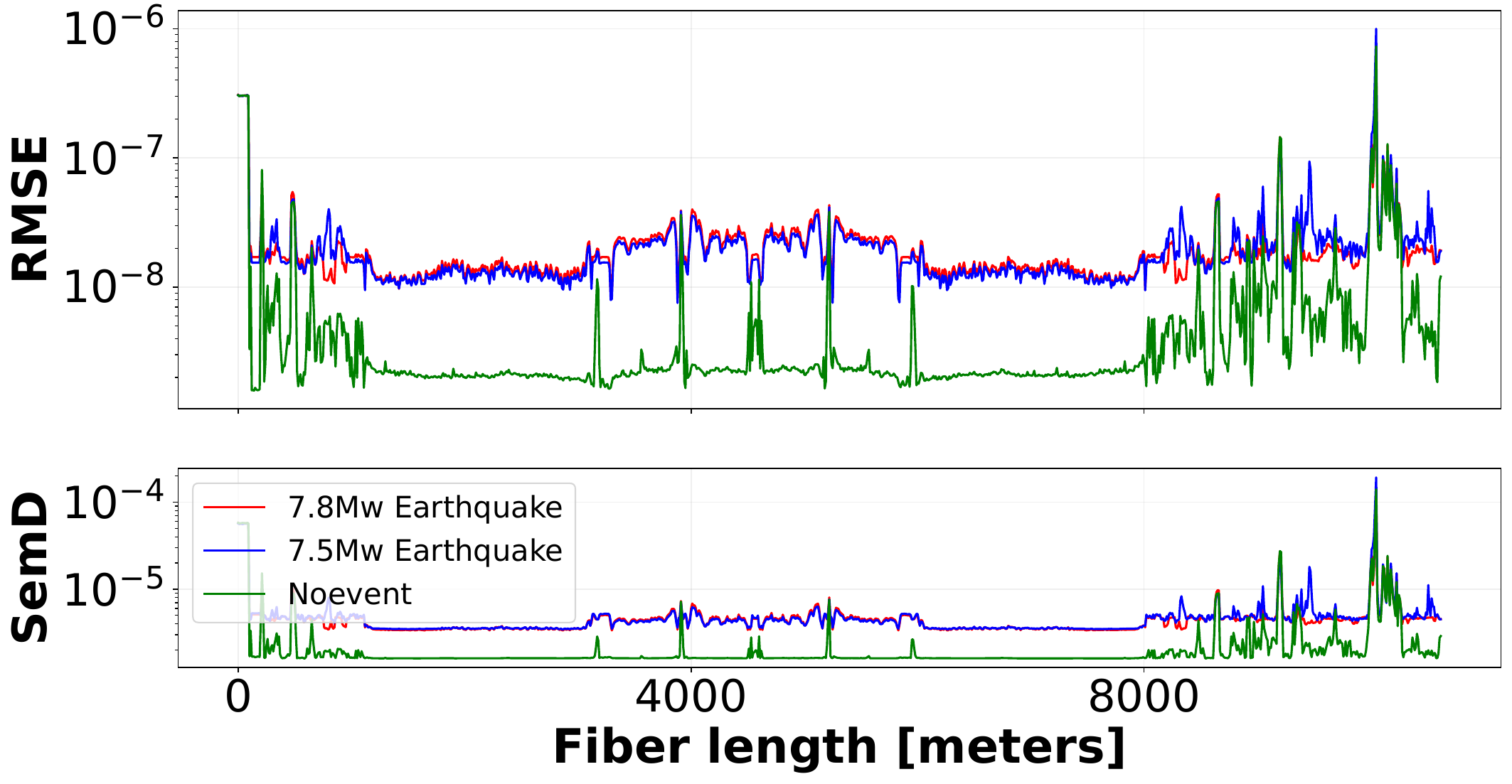}}
	\caption{RMSE comparison between earthquake and no-event periods. During earthquakes, RMSE values are on the order of $10^{-8}$, while quiet conditions show a reduction to ~$10^{-9}$ (one order of magnitude lower), demonstrating sensitivity to coherent seismic energy. Even in noisier sections, earthquake RMSE values remain above the no-event baseline, indicating that the metric captures signal related energy rather than random noise.\label{fig61}}
\end{figure}

A comparison of the two seismic events reveals a narrow but measurable difference in the quietest fiber section. During the Mw 7.8 earthquake, the SemD in the underground region is approximately 0.1 dB lower than during the Mw 7.5 event. Although small, this difference suggests slight variations in effective coupling or environmental conditions between the two recordings. Notably, the Mw 7.8 earthquake occurred  approximately 9 hours earlier than the Mw 7.5 earthquake which occurred approximately at 10:24 UTC. Given the known temperature sensitivity of optical fibers, diurnal temperature variations may have influenced the fiber ground coupling conditions. The observed difference in SemD  therefore likely reflects small thermally induced changes in coupling and strain transfer characteristics, which are detected by the algorithm.

When comparing earthquake recordings to the no-event dataset acquired the following day, a clear difference emerges. During the no-event period, the SemD  in the quiet underground section is approximately -58 dB, corresponding to a reduction of about 3 dB relative to the earthquake cases.
This behavior can be interpreted as follows: during no-event periods, the DAS system predominantly records background and instrument noise. In this regime: signal amplitudes are low,
 noise contributions are statistically similar across neighboring channels if dominated by instrument or self-noise,
spatial coherence is weak but amplitude differences remain small, as a result, the inter-channel differences are minimal, leading to lower semblance distance values. In essence, the algorithm is measuring the effective noise floor of the array. In contrast, during an earthquake where coherent seismic waves propagate along the fiber, local geological conditions and channel-dependent coupling modify the recorded DAS amplitudes and phases and the directional sensitivity of DAS also leads to amplitude variations between neighboring channels. Although the channels remain largely coherent, they are no longer identical, resulting in a modest increase in semblance distance.
\vspace{0.2cm}

\noindent {\large \textbf{\textit{Validating using RMSE}}}

\noindent To validate the performance of the semblance distance method, the RMSE was computed for each channel, as shown in Fig. \ref{fig61}. The RMSE profiles obtained for the Mw 7.8 and Mw 7.5 earthquakes show spatial consistency and nearly identical shapes. This similarity shows that the algorithm produces stable and repeatable results, captures intrinsic array characteristics independent of the specific seismic excitation. 

During the earthquake recordings, the RMSE values are on the order of $10^{-8}$, whereas during the no-event period they decrease to approximately $10^{-9}$, i.e., one order of magnitude lower. This reduction during quiet conditions shows that the metric is sensitive to coherent seismic energy. Even in sections exhibiting elevated noise, the earthquake RMSE values remain  above the no-event baseline, indicating that the metric is able to distinguish between signal related energy and the random noise fluctuations.

Spatially, the fiber sections between 0 and 1100 m and between 8000 and 12,500 m display spikes, amplitude increases, and unstable RMSE behavior. In contrast, the central section between approximately 1200 m and 8000 m shows a structured and spatially coherent pattern that is reproducible across all datasets. These spike features correspond to physical variations along the fiber, including changes in coupling conditions, geological fiber transitions, or fiber quality variations.

\vspace{0.5cm}

\noindent{\large \textbf{\textit{Validating using SNR}}}

\begin{figure}
	\centerline{\includegraphics[width=3.5in]{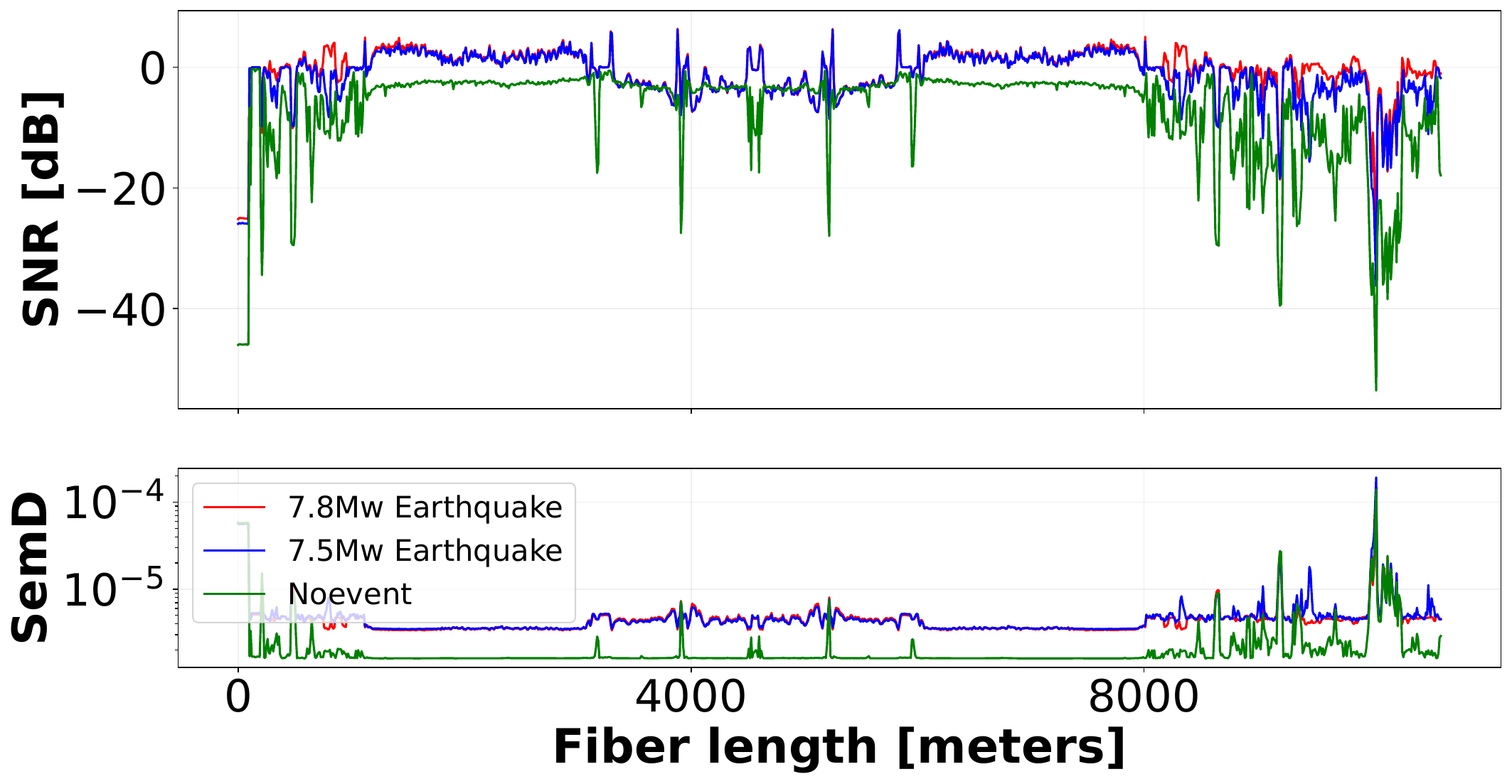}}
	\caption{SNR along the fiber for earthquake and no-event periods. In the central section (from 1100 to 8000 m), SNR remains positive (between 0 and 6 dB) for both events, with nearly identical profiles, indicating stable array response and good coupling (e.g., 6.3-6.4 dB at 3880 m). During quiet conditions, SNR drops slightly below 0 dB, reflecting a stable noise floor. In contrast, the distal sections (0 to 1100 m and 8000 to 12,500 m) show fluctuating and negative SNR, reaching down to -50 dB, indicative of higher noise and poorer coupling.\label{fig62}}
\end{figure}

\noindent To metric output was also compared with the computed SNR. The SNR was calculated in decibels (dB) for every channel in the array, as shown in Fig. \ref{fig62}. In the central portion of the fiber (between 1200 and 8000 m), SNR values during both earthquakes are predominantly positive and range between 0 and 6 dB. The spatial profiles of the Mw 7.8 and Mw 7.5 events are nearly identical, with differences typically below 0.5 dB, confirming the stability of the array response. For example, at 3880 m; one of the quietest channels in the network, the SNR reaches 6.4 dB during the Mw 7.8 earthquake and 6.3 dB during the Mw 7.5 event (a 0.1 dB difference due to coupling or environmental related changes). During the no-event period, the same region exhibits slightly negative SNR values (typically -3 to -6 dB), reflecting a stable noise floor in the absence of coherent excitation. This behavior is characteristic of well coupled, low-noise DAS channels.

In contrast, the distal sections of the fiber (0 to 1100 m and 8000 to 12,500 m) exhibit different behavior. During the earthquakes, the SNR fluctuates and frequently drops into negative values. During the no-event period, these sections record negative and unstable SNR values, ranging from -20 dB down to -50 dB. For instance, at 10,050 m; one of the noisiest channels in the network, the SNR is -33.1 dB during the Mw 7.8 earthquake and -36.2 dB during the Mw 7.5 event (around 3 dB difference due to coupling or environmental related changes). Noise power in these channels exceed signal power by more than three orders of magnitude.

The spatial coincidence between degraded SNR and elevated, unstable RMSE confirms that these regions are dominated by poor coupling, installation artifacts, or strong environmental noise sources which is equally confirmed by the semblance distance algorithm.

\subsection{CONTROLLED STONE DROP EXPERIMENT}

\begin{figure}
	\centerline{\includegraphics[width=3.5in]{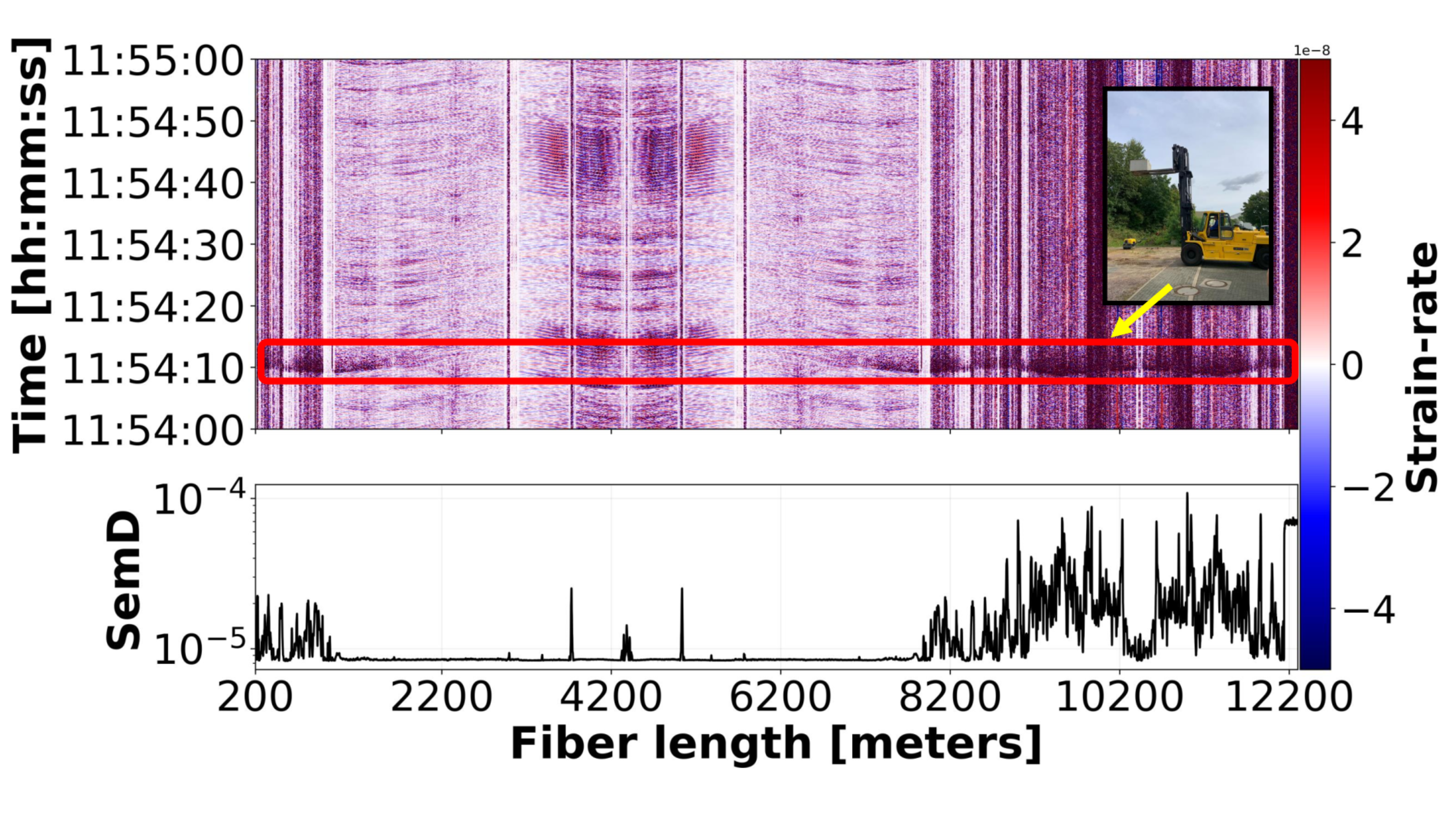}}
	\caption{Waterfall plot and semblance-distance analysis from a controlled seismic source experiment (stone-drop test) conducted in September 2023 on the research campus. A 1- tonne concrete cube was lifted to 5 m using a forklift and released, generating a localized broadband impulsive signal. This experiment complements prior distant earthquake observations, with the semblance-distance algorithm revealing a spatial pattern consistent with those earlier results.\label{fig67}}
\end{figure}

To further evaluate the capability of the channel quality assessment metric, we applied the same methodology to a controlled seismic source experiment conducted in September 2023 on the research campus. The experiment consisted of lifting a 1 tonne concrete cube to a height of approximately 5 m using a forklift and subsequently releasing it, thereby generating a localized broadband impulsive excitation. The resulting ground vibrations were recorded by the DAS system along the same 12.5 km fiber optic network used for the earthquake analysis. This experiment provides a validation scenario because it represents a local and controlled seismic source, in contrast to the distant large-magnitude earthquakes analyzed previously. Although the seismic energy generated by the stone drop is smaller than that of the Mw 7.8 and Mw 7.5 events, the controlled nature of the source enables a direct assessment of whether the algorithm identifies noisy and well coupled channels independent of excitation type and energy level.

Application of the semblance distance algorithm reveals a spatial pattern that is consistent with the earthquake-based results as shown in Fig. \ref{fig67}. The fiber sections from approximately 200 to 1100 m and from 8000 to 12,500 m exhibit elevated semblance distance values, approaching 
$10^{-4}$
, indicating high noise levels and poor coupling conditions. In contrast, the central portion of the array, between approximately 1100 m and 8000 m, displays lower semblance distance values (below $10^{-5}$), corresponding to the quietest and best coupled channels. As observed in the earthquake analysis, localized spikes are present within this otherwise stable region. 

\vspace{0.2cm}

\noindent {\large \textbf{\textit{Validation}}}

\noindent To validate these findings, we computed the 
SNR along the fiber for the stone-drop event. The SNR profile shows clear spatial variability consistent with the semblance distance results. The central section (approximately from 1100 to 8000 m) shows relatively stable SNR values, typically ranging between -3 dB and -8 dB. Although lower than the positive SNR values observed during earthquake events, these values remain spatially consistent and indicate that the impulsive signal remains detectable above the background noise level.

In contrast, the sections between 200 (0 to 200m cropped) and 1100 m and beyond 8000 m show SNR degradation. In these regions, SNR values fall below -20 dB and in some cases approach -40 dB, indicating that the noise power exceeds signal power. Such behavior suggests  environmental contamination, poor mechanical coupling, or installation related artifacts.

An inverse relationship is observed between the SNR and semblance distance metrics. Regions with higher SNR values correspond to low semblance distance values. Conversely, sections characterized by negative SNR values exhibit large  variable semblance distance values, reflecting a breakdown of spatial coherence due to noise. The spatial agreement between these two independent validation metrics provides evidence that the proposed algorithm identifies noisy and poorly coupled fiber sections.

\begin{figure}
	\centerline{\includegraphics[width=3.5in]{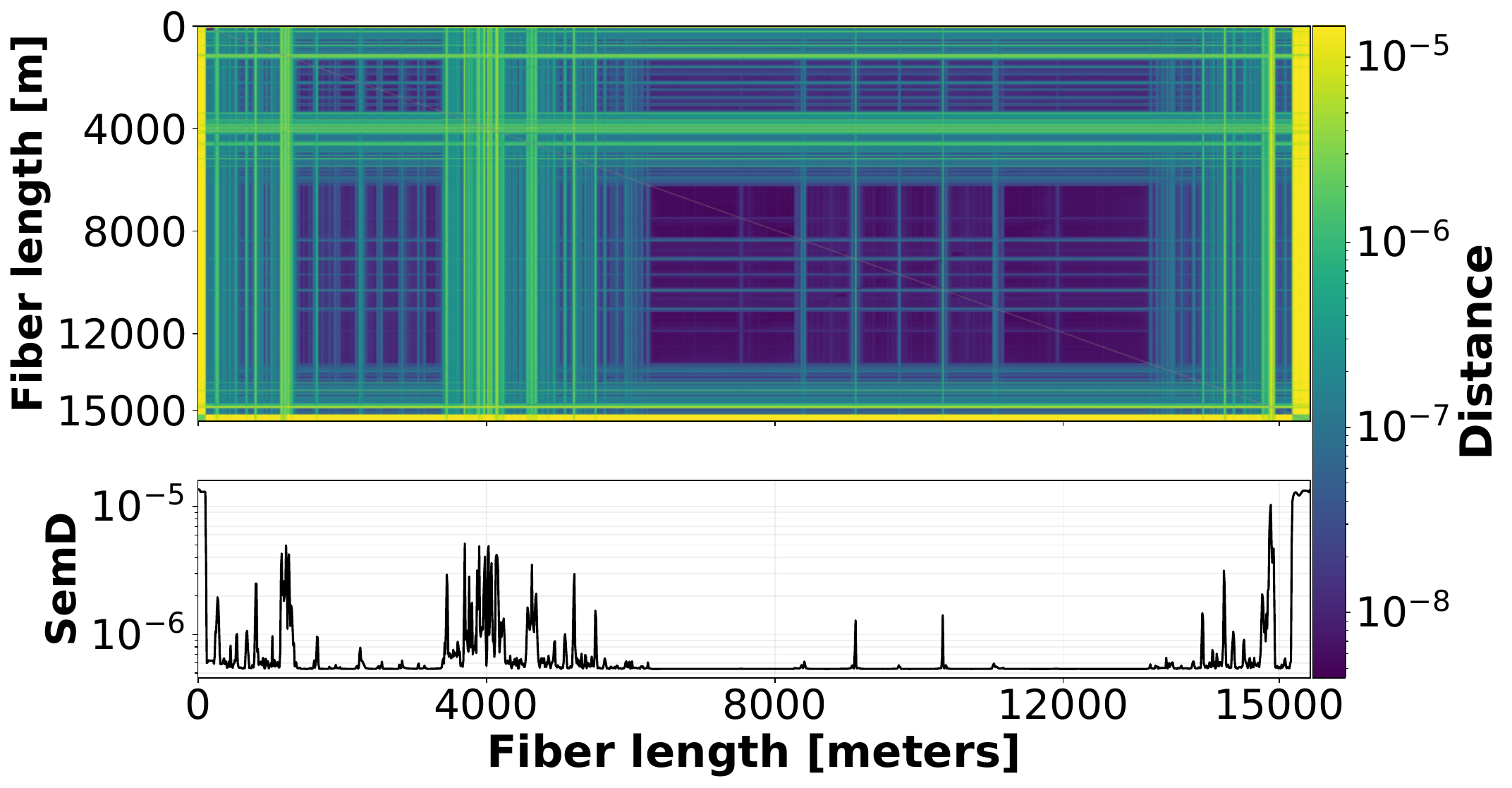}}
	\caption{Extension of the dark-fiber network from 12.5 km to 15 km in 2024 enabled evaluation of the algorithm under updated geometry and installation conditions. Semblance distance analysis identifies the 6200 to 13,000 m segment as the lowest noise, best coupled region, with strong spatial coherence ($10^{-8}$). The newly added 1500 to 3500 m section, routed through tunnels, shows similarly low noise despite not being deeply buried, while the 3500 to 5300 m campus exposed to anthropogenic activity exhibits higher semblance distances.\label{fig8}}
\end{figure}


\subsection{NOISY CHANNELS IN THE FIBER NETWORK 2025}
In 2024, the dark fiber network was extended to a total length of 15 km. We applied the same semblance distance based analysis to the extended array configuration.
The results  indicate that the channels between approximately 6200 m and 13,000 m represent the lowest noise and best coupled portion of the 15 km network as shown in Fig. \ref{fig8}. In this region, the semblance distance reaches values on the order of 
$10^{-8}$
, reflecting low inter-channel dissimilarity. Notably, the newly added fiber segment spans approximately 1500 to 3500 m. Although this section is not buried at a depth of 20 m like the previously identified underground portion, it is routed through tunnel infrastructure and is therefore shielded from direct exposure to surface level anthropogenic disturbances. Consistent with this installation condition, the tunnel based segment shows similar low semblance distance values, comparable to those observed in the 6200 to 13,000 m region.

In contrast, the fiber section between approximately 3500 m and 5300 m, located within the campus region and exposed to significant anthropogenic activity, displays elevated distance values. This section is subject to multiple environmental noise sources, including vehicular traffic, mechanical infrastructure, and surface vibrations. The algorithm differentiates this noisy segment from the quieter tunnel and underground sections, demonstrating its ability to adapt to the modified network configuration. 

Within both from 6200 to 13,000 m and from 1500 to 3500 m regions, localized spikes are observed in the distance metric. These spikes correspond to specific routing transitions where the fiber is temporarily exposed to the external environment. Due to the high strain sensitivity of DAS, even minor exposure to ambient disturbances results in measurable increases in inter-channel dissimilarity. The presence of these spikes further confirms that the algorithm detects true installation dependent variations rather than artificial processing artifacts.

\subsection{TEMPORAL EVOLUTION OF COUPLING CONDITIONS}
\begin{figure}
	\centerline{\includegraphics[width=3.5in]{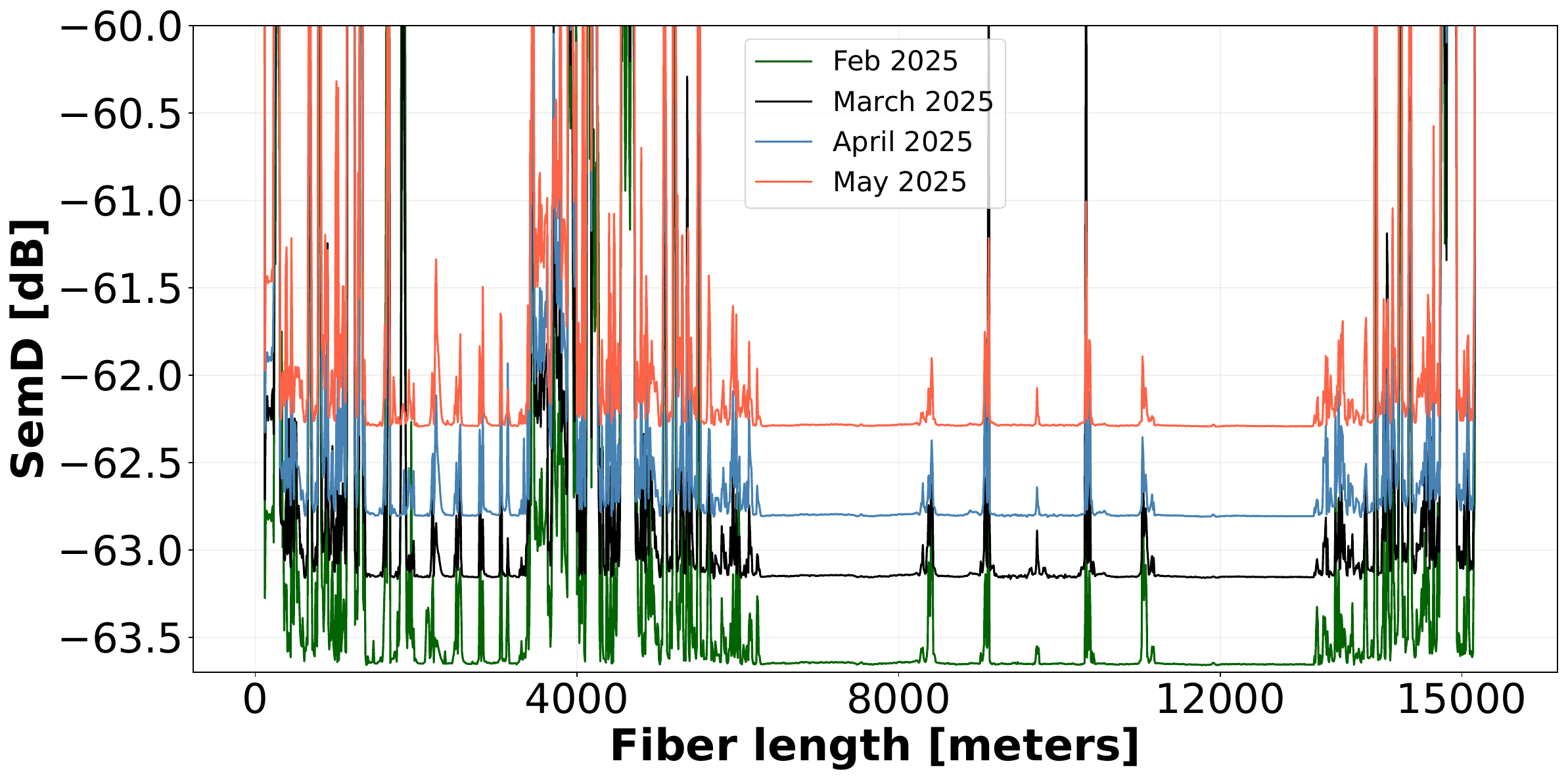}}
	\caption{Temporal stability of the DAS array assessed using ultra low frequency data collected from February to May. Monthly semblance distance profiles reveal stable spatial patterns alongside a gradual increase in average distance, from ~-63.6 dB in February to ~-62.3 dB in May. This steady rise (~between 0.3 and 0.5 dB per month) indicates a slow change in coupling conditions rather than random variation.\label{fig9}}
\end{figure}

To investigate the temporal stability of the DAS array, we analyzed ultra low-frequency data acquired over a four month period between February and May. At such low sampling frequencies, the DAS system is sensitive to ultra low-frequency strain phenomena, including thermal strain, cable relaxation, and slow ground deformation. This regime provides insight into long-term fiber ground coupling behavior rather than transient seismic events.
The monthly distance profiles show both persistent spatial features and systematic temporal variations. February exhibits the lowest average distance (approximately -63.6 dB), followed by a gradual increase in March (0.4 dB higher), a further increase in April (0.4 dB additional), and the highest values in May (approximately -62.3 dB) as shown in Fig. \ref{fig9}. The monotonic progression of roughly between 0.3 to 0.5 dB per month suggests a slow and global structural change in coupling conditions rather than random fluctuations. This shift occurs uniformly along most of the fiber length, indicating a global environmental influence rather than isolated disturbances.

\begin{figure}
	\centerline{\includegraphics[width=3.5in]{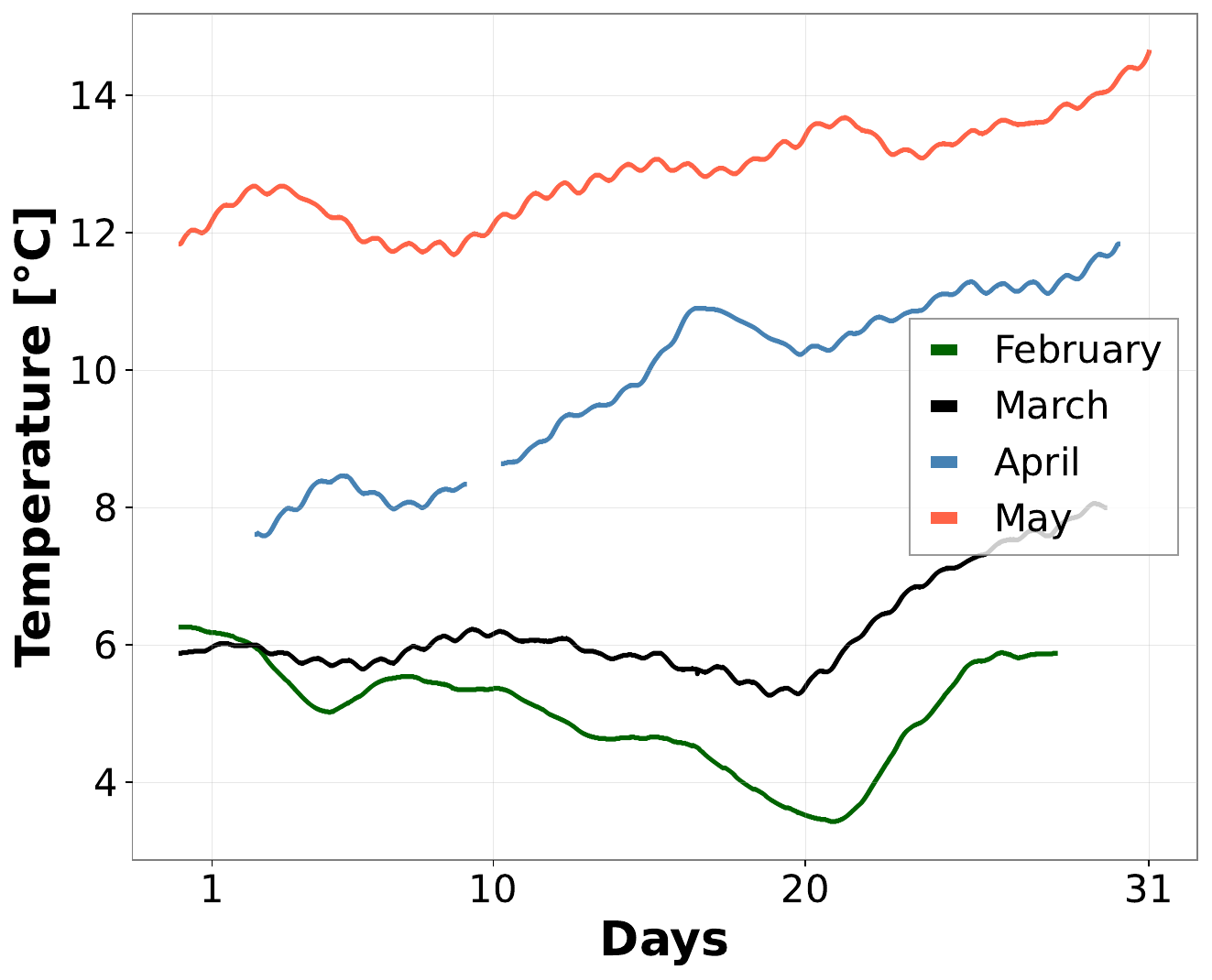}}
	\caption{Soil temperature measured in Hamburg at 0.8 m depth during the four-month study period in 2025. Monthly mean temperatures follow the expected seasonal progression: February is the coldest (5.13 °C), followed by March (5.88 °C), April (10.34 °C), and May (12.92 °C). The burial-depth measurement reflects the sustained thermal state at the fiber location and provides independent ground truth for the seasonal coupling trend identified by the semblance distance metric (Section V.D).\label{fig99}}
\end{figure}

The observed trend is consistent with seasonal temperature variations in Hamburg, where February represents the coldest period and May corresponds to a  warmer early summer as shown in \ref{fig99}. This trend suggests that temperature acts as an amplifying factor, as temperatures increase from late winter to spring, thermal expansion of both soil and cable materials can modify burial contact conditions and cable tension. Potential mechanisms include soil expansion, reduced ground stiffness, changes in moisture content, and small variations in cable-soil contact pressure. These processes would reduce mechanical coupling stiffness, thereby increasing inter-channel dissimilarity and raising the distance metric.

In addition to the global baseline drift, spatial anomalies are observed at approximately from 3500 to 4500 m, some peaks between 8000 and 9000 m, and anomalies near 14,500 m. These features appear at identical spatial locations in all four monthly datasets, suggesting that they correspond to permanent infrastructure characteristics such as road crossings, fiber splices, conduit transitions, or cable bends. Although their positions remain fixed, their amplitudes increase slightly in later months, particularly in May. This behavior indicates that existing coupling irregularities become more pronounced under changing environmental conditions.

The ability to detect such subtle, sub-decibel changes shows the capability of the proposed algorithm. Without the semblance distance metric, gradual 0.3 to 0.4 dB step changes would be  difficult to identify through visual inspection of raw DAS strain data. The algorithm can effectively functions as a fiber health and coupling monitoring tool, capable of detecting slow mechanical and environmental changes along the sensing cable.

\subsection{COMPARISON OF SEMBLANCE DISTANCE METHOD TO CONVENTIONAL METHODS}
The low-frequency DAS data acquired over a four-month period was used to compare the semblance distance method's performance against that of three conventional approaches: SNR, RMS, and correlation or coherence based metrics. The goal was to demonstrate the inability of conventional approaches to detect and resolve subtle, low-frequency, spatially distributed variability in DAS measurements. Entropy and spatial gradient energy were used as additional metrics to evaluate the observed trends.

The SNR was computed using a smoothing based approach. For each sensor and each month, the time series was smoothed using a 4 hour moving window to estimate the signal component. The noise was defined as the residual between the raw and smoothed data. The SNR was then calculated as the ratio of signal variance to noise variance using the conventional definition. For the correlation based method, an ensemble average across all DAS channels was first computed, assuming spatial homogeneity within the WAVE network. This ensemble average was then correlated with each individual sensor time series to produce a correlation based measure. Spatial gradient energy was computed using the same temporal window, capturing localized spatial transitions. Entropy was used to quantify overall variability and randomness in the signal.

\begin{figure}
	\centerline{\includegraphics[width=3.5in]{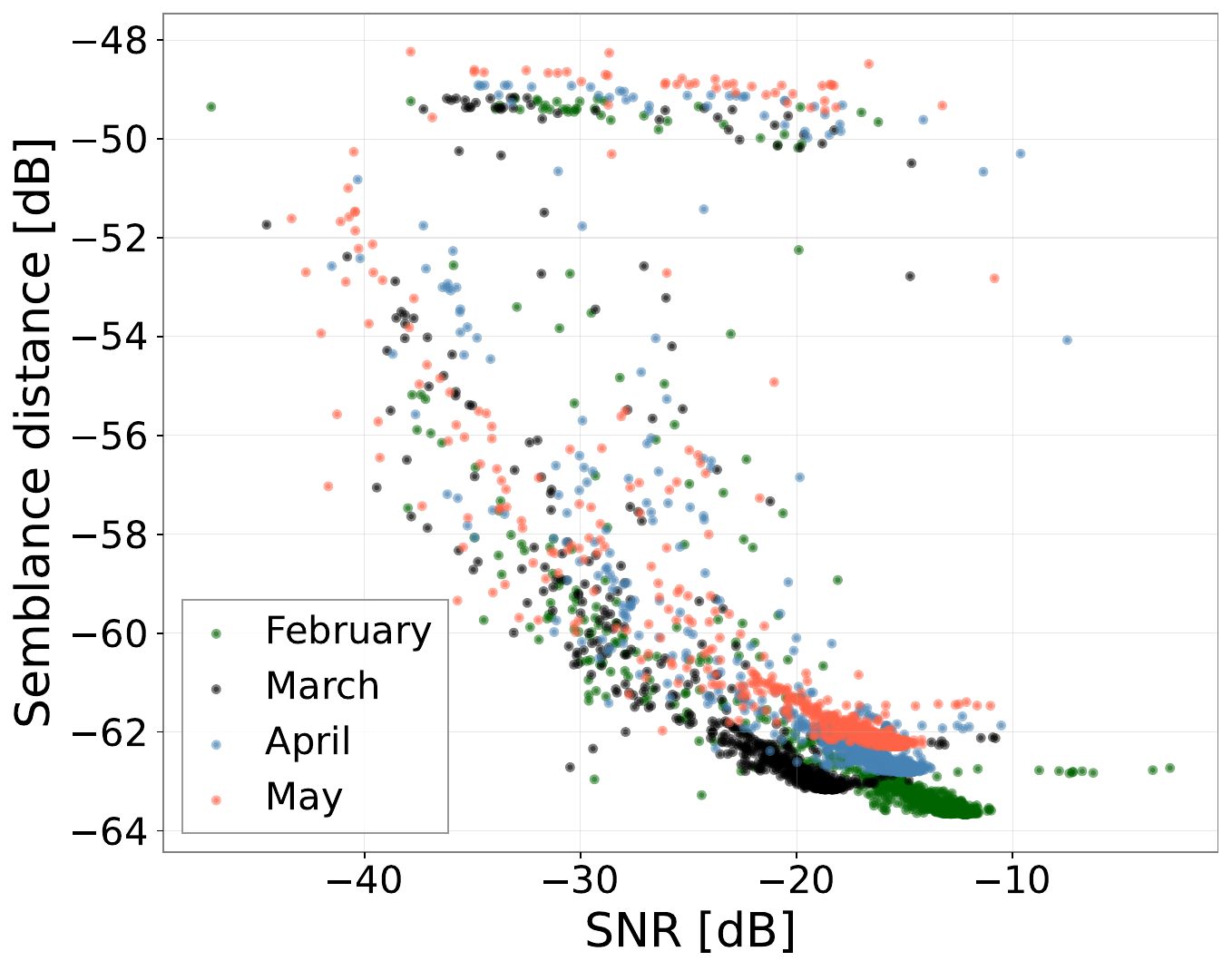}}
	\caption{The channel-wise relationship between SNR and semblance distance is shown across the four-month low-frequency dataset. The two metrics are negatively correlated every month (Pearson's $r$ = -0.77 to -0.73, $p$ $<$ 0.0001), which confirms that channels that are identified as low-SNR using the conventional metric are also identified as high-distance using the proposed method.\label{fig2080}}
\end{figure}

We also quantified the relationship between SNR and SemD directly at the channel level. For each month, Pearson correlation was computed between the SNR and SemD values of all channels as shown in Fig. \ref{fig2080}. All four months show a strong negative correlation ($r$ $=$ -0.77, -0.76, -0.74, and -0.73 for February, March, April, and May respectively; $p$ $<$0.0001 in all cases), confirming that the two independently computed metrics agree on relative channel quality across the range of coupling conditions present in the network. Notably, the correlation strength decreases slightly but monotonically from February to May. This trend mirrors the seasonal coupling degradation discussed in Section V.D and suggests that as ground coupling weakens during the spring, the agreement between SNR and SemD loosens slightly, possibly because SNR estimation becomes less stable under degraded coupling conditions, while SemD, which does not rely on an explicit noise model, remains more robust. 

The well-coupled channels (SNR above approximately -20 dB) form a tight, low-scatter cluster in which the four months remain cleanly separated and monotonically ordered by season, consistent with the sub-decibel seasonal drift identified. In contrast, the poorly coupled channels show substantially greater scatter and inter-month overlap,  reflecting the heterogeneous and less predictable noise environment in those fiber sections.

\begin{figure}
	\centerline{\includegraphics[width=3.5in]{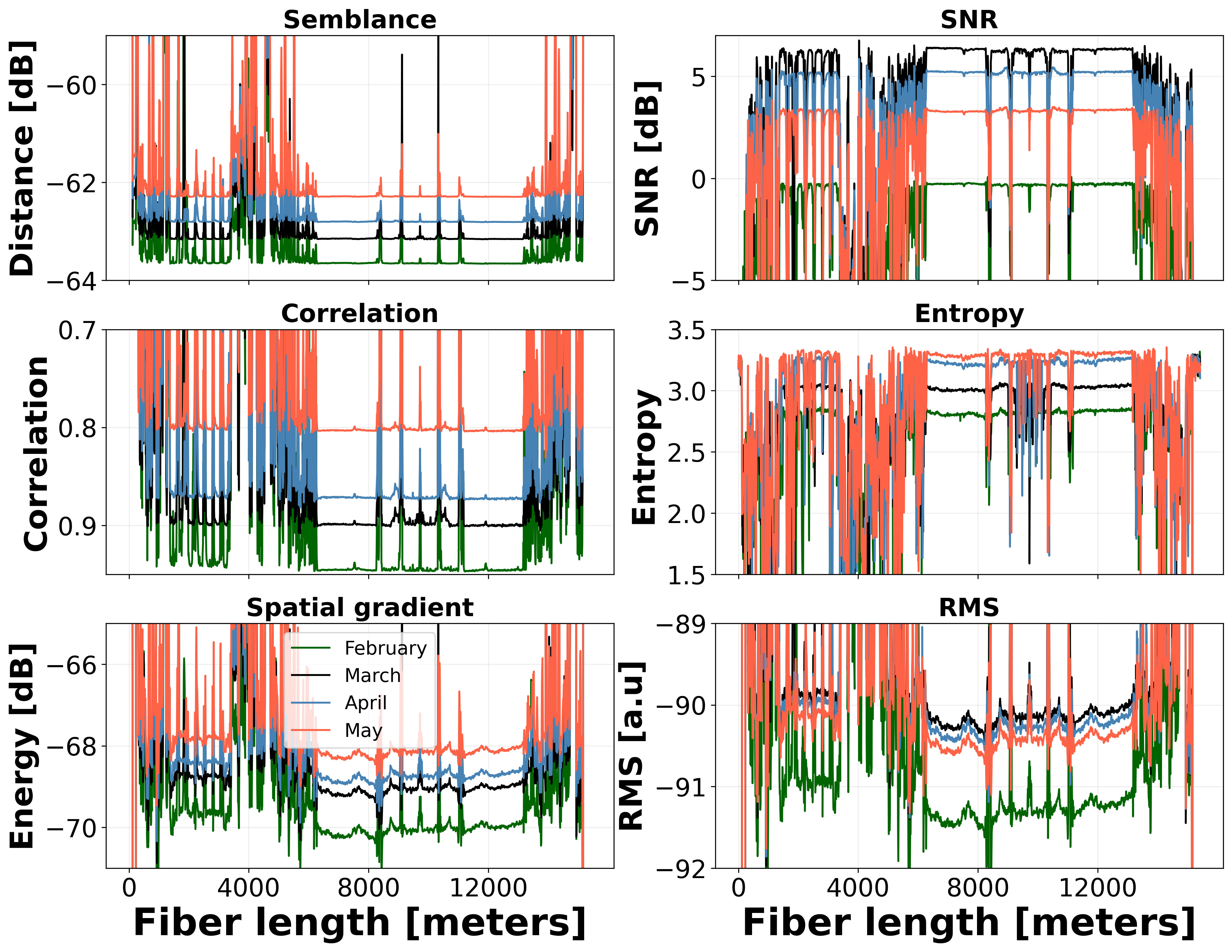}}
	\caption{Semblance distance captures subtle DAS variations over four months, resolving small differences (0.3–0.5 dB) between consecutive months. In contrast, SNR and RMS misrepresent the temperature trend, incorrectly ranking months and contradicting entropy and spatial gradient energy analyses, which follow the expected progression from February (coldest) to May (warmest).\label{fig200}}
\end{figure}

In a one-to-one comparison, the SemD method is more sensitive to local extremes than the SNR and RMS based metrics. For instance, within the range of 6,000 to 13,500 meters; the quietest segment of the WAVE network, all three methods appear visually flat (see Fig. \ref{fig200}). However, the semblance metric reveals constant microoscillations across the months in the form of small, periodic spikes. These features indicate structural variations along the fiber, such as installation constraints or heterogeneous soil segments. This interpretation is supported by the fact that these oscillations occur at consistent spatial locations across all four months while gradually increasing in amplitude from February to May. SNR and RMS metrics smooth and average out these variations or classify them as noise due to their reliance on temporal smoothing and averaging. While these methods provide stable, noise reduced estimates of overall signal levels, they suppress localized fluctuations.

\begin{table}[!t]
	\caption{Comparison of Monthly Rankings: True Hamburg Mean Soil Temperature vs. Channel Quality Metrics (1=coldest, 4=warmest)}
	\label{tab:rank_comparison}
	\centering
	\footnotesize
	\begin{tabular}{lcccc}
		\toprule
		& Feb & Mar & Apr & May \\
		\midrule
		True Soil Temp (\textdegree C) & 5.13 & 5.88 & 10.34 & 12.92 \\
		True Rank                 & 1    & 2    & 3    & 4 \\
		SemD Rank                 & 1    & 2    & 3    & 4 \\
		SNR Rank                  & 1   & 4    & 3   & 2 \\
		RMS Rank                  & 1   & 4    & 3    & 2 \\
		Entropy Rank                  & 1   & 2    & 3    & 4 \\
		Spatial Gradient Rank                  & 1   & 2    & 3    & 4 \\
		\bottomrule
	\end{tabular}
	
	\vspace{4pt}
	\footnotesize Spearman $\rho$ vs.\ true rank: SemD = 1.00, SNR = 0.20, RMS = 0.20
\end{table}

Furthermore, the method can detect subtle variations. In quantitative terms, it can detect differences of approximately 0.4 dB between February and March, 0.3 dB between March and April, and 0.5 dB between April and May. However, both the SNR and RMS metrics fail to reproduce the expected temperature trend. While it correctly identifies February as the coldest month with the lowest SNR, it incorrectly ranks May as the second coldest, April as the third coldest and March as the warmest. A similar inconsistency is observed with the RMS metric. These results contradict auxiliary analyses, including entropy and spatial gradient energy, which align with the expected temperature trend in Hamburg. This discrepancy highlights the limitations of the SNR and RMS metrics, which are based on assumptions that suppress localised spatial variations.

Table II compares the monthly ranking implied by each metric against the true ranking from independently measured mean soil temperature in Hamburg at a depth of 0.8m over the same period (Feb: 5.13°C, Mar: 5.88°C, Apr: 10.34°C, May: 12.92°C). SemD reproduces the true temperature ranking exactly  (Spearman $\rho$ = 1.00), while SNR ranking and RMS shows only a weak positive association ($\rho$ = 0.20), complementing the much larger-sample channel-level correlation reported in Fig. \ref{fig2080}.

The observed temperature-dependent variations from February to May are non-linear and spatially heterogeneous, reflecting changes in fiber ground coupling rather than simple amplitude scaling. This suggests that the system has nonlinear thermal sensitivity which the semblance method effectively captures; however, SNR and RMS, being power-based measures, are insensitive to such effects and instead average them out.

A key advantage of the semblance approach is that it preserves phase-aligned spatial coherence: peaks and troughs remain spatially aligned across months, indicating that the observed variability is structured and repeatable rather than stochastic, and suggesting that the signal contains persistent spatial features modulated by environmental conditions. Other metrics exhibit significant limitations. Correlation-based methods, for example, capture overall similarity but tend to obscure small deviations, entropy quantifies global randomness without distinguishing coherent spatial structures, and spatial gradient energy highlights abrupt changes but fails to capture slow baseline drifts. A further advantage of semblance is that it provides inherent denoising through coherence detection, distinguishing signal from noise in low-noise regions without the need for explicit filtering; by comparison, SNR and RMS rely on assumptions that reduce sensitivity to subtle spatial variability, while correlation and entropy lose interpretability under noisy conditions.

\section{APPLICATIONS}
The algorithm can be applied across a wide range of engineering and scientific applications. One application lies in structural health monitoring. DAS systems are increasingly deployed along civil infrastructure such as bridges, tunnels, aircraft surfaces, pipelines, and high-rise buildings. Slow mechanical changes occurring over days or longer time scales can produce subtle strain variations along the sensing fiber. The  algorithm can help detect and localize these irregularities by identifying channel-specific deviations. In particular, small cracks, developing faults, or structural degradation may manifest as irregular patterns across neighboring DAS channels. By identifying such anomalies, the method can lead to early detection of structural damage that might otherwise remain undetected in conventional measurements and processing.

 Another potential application is in future planetary exploration missions. Several proposed lunar and planetary missions plan to deploy DAS systems on the surface of the Moon or other celestial bodies. Due to the sensitivity of DAS systems, these fibers are likely to record a wide variety of signals originating from environmental processes, surface activity, and potential seismic events. The algorithm could assist in identifying channels with high signal quality and detecting channel-specific events that may appear incoherent but still contain valuable information for planetary seismology. The algorithm may help improve the interpretation of weak seismic signals and enhance the detection of subtle lunar or planetary surface processes.
 
 The method is also relevant for high sensitivity research facilities such as gravitational wave observatories. Modern gravitational wave detectors rely on networks of environmental sensors to measure and subtract noise sources, including seismic and Newtonian noise. The effectiveness of such noise cancellation strategies depends on selecting auxiliary channels with high SNR while excluding channels dominated by noise. The algorithm can assist in identifying the most informative DAS channels for environmental monitoring and noise subtraction, thereby improving the performance of sensor based noise cancellation systems. 

Finally, the method may also benefit seismic imaging and subsurface characterization studies. In these applications, high quality data are essential for achieving accurate imaging results. The algorithm can assist in identifying high SNR channels and excluding unreliable measurements, thereby improving the overall quality of seismic imaging and inversion results.

\section{CONCLUSION}
In this study, we introduced a statistical framework for identifying low-noise and poorly coupled channels in DAS arrays using a semblance distance based analysis. The proposed method leverages the spatial coherence of signals recorded by neighboring fiber channels to quantify inter-channel dissimilarity and thereby distinguish between well coupled, high quality sensing segments and sections dominated by noise or poor mechanical coupling. The algorithm was validated using independent datasets recorded by a campus scale DAS network. These datasets included two earthquake events, specifically the 2023 Mw 7.8 and Mw 7.5 Turkish-Syrian earthquakes, a controlled human-induced excitation in the form of a stone-drop experiment, and long duration background recordings obtained during quiet periods. Across all scenarios, the algorithm identified the same spatial regions of the fiber as either low-noise or noise dominated.

Further development of the semblance distance algorithm can be done. For example, evaluating the performance of the algorithm across different frequency bands. While the present study primarily focused on frequencies from the sub-Hz range up to approximately 48 Hz, future work could investigate the algorithm’s behavior at higher frequencies where DAS systems capture additional environmental and anthropogenic signals. The semblance-distance algorithm can be modified to detect irregular or transient channel behavior in real time. Incorporating temporal analysis of the distance metric could allow the algorithm to identify the precise timing of anomalous events or sudden coupling changes along the fiber. Such capabilities would be valuable for real time monitoring of DAS network performance and rapid detection of installation related issues. Further research could also explore integration of the algorithm with machine learning. Combining channel quality metrics with machine learning may enable automated classification of fiber conditions, improved noise suppression strategies, and adaptive channel selection for large scale DAS arrays.

\section*{ACKNOWLEDGMENT}
This research was supported in part through the Maxwell computational resources operated at Deutsches Elektronen-Synchrotron DESY, Hamburg, Germany.

\section*{REFERENCES}

\def\refname{\vadjust{\vspace*{-1em}}} 


\begin{thebibliography}{00}
	
\bibitem{b9}  H.F. Wang, X. Zeng, D.E. Miller, D. Fratta, K.L. Feigl, C.H. Thurber, and R.J. Mellors,“Ground motion response to an ML 4.3 earthquake using co-located distributed acoustic
sensing and seismometer arrays”, Geophys. J. Int., 213, 2020-2036, 2018.


\bibitem{b10} A. H. Hartog, An Introduction to Distributed Optical Fibre Sensors, 1st ed. Boca Raton, FL, USA: pp: 3-103, CRC Press, 2017 . doi: 10.1201/9781315119014.


\bibitem{t0} J. Mestayer, B. Cox, P. Wills, D. Kiyashchenko, J. Lopez,
and M. Costello, “Field trials of distributed acoustic sensing
for geophysical monitoring”, In: 81 st SEG Annual International Meeting, Expanded, San Antonio, USA, 2011, pp.
4253-4257.	


\bibitem{t1} A. Mateeva, J. Mestayer, B. Cox, D. Kiyashchenko, P. Wills,
and S. Grandi, “Distributed Acoustic Sensing (DAS) for reservoir monitoring with VSP”, In: 2 nd EAGE Borehole Geo-
physics Workshop - 3D VSP: Benefits, Challenges and Po-
tential, St. Julain's, Malta, 2013.


\bibitem{t2} M. Li, H. Wang, and G. Tao, “Current and future applications of distributed acoustic sensing as a new reservoir geophysics tool,” The Open Petroleum Engineering Journal, vol. 8, Jun. 2015, doi: 10.2174/1874834120150625E008.


\bibitem{t3} T.M. Daley, B. M. Freifeld, J. Franklin, S. Dou, R. Pevzner,
V. Shulakova, S. Kashikar, D. E. Miller, J. Goetz, J. Henninges, and S. Lueth, “Field testing of fiber-optic distributed
acoustic sensing (DAS) for subsurface seismic monitoring”,
The Leading Edge, vol. 32, no. 6, pp. 699-706, 2013.
	
	
	
\bibitem{a1} Y. Li, M. Karrenbach, and J. B. Ajo-Franklin, “A literature review: distributed acoustic sensing (DAS) geophysical applications over the past 20 years,” in Distributed Acoustic Sensing in Geophysics: Methods and Applications, pp. 229-291, 2021.

\bibitem{a0} Z. H. Zhu, W. Liu, T. Wang, J. W. Su, and B. Shi, “Distributed acoustic sensing for monitoring linear infrastructures: Current status and trends,” Sensors, vol. 22, no. 19, p. 7550, Oct. 2022, doi: 10.3390/s22197550.

\bibitem{a00} R. Di Sante,  “Fibre optic sensors for structural health monitoring of aircraft composite structures:
Recent advances and applications,”. Sensors, 15, 18666-18713, 2015.

\bibitem{b1} O. Bölt, L. Cristiano, S. Croatto, D. Gajewski, E. Genthe, O. Gerberding, C. Hadziioannou, C. Hammer, M. Hoffmann, K.-S. Isleif, A. Kiel, C. M. Krawczyk, R. Maass, I. Barbosa, N. Meyners, R. Rading, H. Schlarb, R. Schnabel, W. Vossius, and C. Wollin, “Distributed acoustic fiber sensing for research campuses and large scientific infrastructures-The Hamburg WAVE proto-network,” \textit{arXiv preprint} arXiv:2511.17141, Nov. 2025.

\bibitem{b2} Q. Zhai, A. Husker, Z. Zhan, E. Biondi, J. Yin, F. Civilini, and L. Costa, “Assessing the feasibility of distributed acoustic sensing (DAS) for moonquake detection,” Earth Planet. Sci. Lett., vol. 635, Art. no. 118695, 2024, doi: 10.1016/j.epsl.2024.118695.




\bibitem{b6} G. Agrawal, "Fiber Optic Communication Systems", A. John Willey and Sons Publications, 4th Edition, 2010.



\bibitem{d1} D. Li,  L. Ren, H. Li, “Mechanical property and strain transferring mechanism in optical fiber sensors,” In Fiber
Optic Sensors, InTech, London, UK, 2012.

\bibitem{d2} H. Wang,  P. Xiang, “Strain transfer analysis of optical fiber based sensors embedded in an asphalt pavement
structure,” Meas. Sci. Technol., 27, 075106, 2016.

\bibitem{d3}  J.M. Henault,  J. Salin, G. Moreau,  M. Quiertant, F. Taillade,  K. Benzarti,  S.Delepine-Lesoille, “Analysis
of the strain transfer mechanism between a truly distributed optical fiber sensor and the surrounding medium,” Proceedings of the Concrete Repair, Rehabilitation and Retrofitting III, Cape Town, South Africa, 2012.	






\bibitem{b4} J. Harms, “Terrestrial gravity fluctuations”, Living Reviews in Relativity, vol. 22, pp. 1-154, 2019.

\bibitem{b5} L. Ehsaninezhad, C. Wollin, V. Rodríguez Tribaldos, B. Schwarz, and C. M. Krawczyk, “Urban subsurface exploration improved by denoising of virtual shot gathers from distributed acoustic sensing ambient noise,” Geophysical Journal International, vol. 237, no. 3, pp. 1751-1764, Jun. 2024, doi: 10.1093/gji/ggae134.



\bibitem{b3} F. Falcetelli, L. Rossi, R. Di Sante, and G. Bolognini, “Strain transfer in surface-bonded optical fiber sensors,” Sensors, vol. 20, no. 11, p. 3100, 2020, doi: 10.3390/s20113100.




\bibitem{e10} N. Wiener, “Generalized harmonic analysis,” Acta Mathematica, vol. 55, no. 1, pp. 117-258, 1930.

\bibitem{e11} C. H. Knapp and G. C. Carter, “The generalized correlation method for estimation of time delay,” IEEE Transactions on Acoustics, Speech, and Signal Processing, vol. ASSP-24, no. 4, pp. 320-327, Aug. 1976.


\bibitem{e20} P. D. Welch, “The use of fast Fourier transform for the estimation of power spectra: A method based on time averaging over short, modified periodograms,” IEEE Transactions on Audio and Electroacoustics, vol. AU-15, no. 2, pp. 70-73, Jun. 1967.

\bibitem{e21} G. C. Carter, C. H. Knapp, and A. H. Nuttall, “Estimation of the magnitude-squared coherence function via overlapped fast Fourier transform processing,” IEEE Transactions on Audio and Electroacoustics, vol. AU-21, no. 4, pp. 337–344, Aug. 1973. 


\bibitem{e30} J. W. Fay, “Confidence bounds for signal-to-noise ratios from magnitude-squared coherence estimates,” IEEE Transactions on Acoustics, Speech, and Signal Processing, vol. ASSP-28, no. 6, pp. 757-758, Dec. 1980.

\bibitem{e31} A. M. F. M. de Sá, “A note on the coherence-based signal-to-noise ratio estimation in systems with periodic inputs,” Journal of the Franklin Institute, vol. 343, no. 7, pp. 688-698, Nov. 2006.






\bibitem{b66} A. Titov, V. Kazei, A. AlDawood, E. Alfataierge, A. Bakulin, and K. Osypov, “Quantification of DAS VSP Quality: SNR vs. Log-Based Metrics,” Sensors, vol. 22, no. 3, p. 1027, 2022, doi: 10.3390/s22031027.

\bibitem{b7} Y. Lu and P. C. Loizou, “Estimators of the magnitude-squared spectrum and methods for incorporating SNR uncertainty,” IEEE Transactions on Audio, Speech, and Language Processing, vol. 19, no. 5, pp. 1123-1137, Jul. 2011.

\bibitem{b8} F. Chen, Y. Kang, H. Yu, and F. Ji,
“Non-data-aided ML SNR estimation for AWGN channels with deterministic interference,”
EURASIP Journal on Wireless Communications and Networking, vol. 2014, no. 45, 2014.

\bibitem{kk1} M. E. Tipping and C. M. Bishop, “Probabilistic principal component analysis,” Journal of the Royal Statistical Society: Series B, vol. 61, no. 3, pp. 611-622, 1999.

\bibitem{kk2} J. Shlens, “A tutorial on principal component analysis,” arXiv preprint arXiv:1404.1100, 2014.

\bibitem{kk3} N. Otsu, “A threshold selection method from gray-level histograms,” IEEE Transactions on Systems, Man, and Cybernetics, vol. 9, no. 1, pp. 62-66, Jan. 1979.


\bibitem{c2} N.S. Neidell and M.T. Taner, “Semblance and other coherency measures for multichannel
Data”, Geophysics, 36, 482-497, 1971.

\bibitem{c1} M.T. Taner, F. Koehler, and R.E. Sheriff, “Complex seismic trace analysis”, Geophysics, 44,
1041-1063, 1979.



\bibitem{c3} T. Shi and S. Huo,“Complex semblance and its application”, J. Earth Sci., 30, 849-852, 2019.


\bibitem{c4} R. Rading, F. Badaracco, S. Beis, K. S. Isleif, P. Ophardt, W. Vossius, and the Wave Collaboration, "Distributed acoustic sensing for environmental monitoring, and Newtonian noise mitigation: Comparable sensitivity to seismometers,"	arXiv:2507.13523, 2025.



\bibitem{c01} C. D. Manning, P. Raghavan, and H. Schütze, Introduction to Information Retrieval. Cambridge, U.K.: Cambridge University Press, 2008.

\bibitem{c02} T. Hastie, R. Tibshirani, and J. Friedman, The Elements of Statistical Learning, 2nd ed. New York, NY, USA: Springer, 2009.


\bibitem{c03} P. C. Mahalanobis, "On the generalized distance in statistics," Proceedings of the National Institute of Sciences of India, vol. 2, no. 1, pp. 49–55, 1936.


\bibitem{c04} C. Villani, Optimal Transport: Old and New. Berlin, Germany: Springer, 2009.

\bibitem{c044} T. M. Cover and J. A. Thomas, Elements of Information Theory, 2nd ed. Hoboken, NJ, USA: Wiley, 2006.


\bibitem{c05} R. Rading and K. Isleif, "Coherent and Incoherent Noise Cancellation Using Distributed Optical Fiber Sensors," 2024 IEEE Photonics Society Summer Topicals Meeting Series (SUM), Bridgetown, Barbados, 2024.






\bibitem{s0} T. K. Vintsyuk, “Speech discrimination by dynamic programming,” Cybernetics, vol. 4, pp. 52-57, 1968.

\bibitem{s1} F. Itakura, “Minimum prediction residual principle applied to speech recognition,” IEEE Transactions on Acoustics, Speech, and Signal Processing, vol. 23, pp. 154-158, 1975.

\bibitem{s2} H. Sakoe and S. Chiba, “Dynamic programming algorithm optimization for spoken word recognition,” IEEE Transactions on Acoustics, Speech, and Signal Processing, vol. 26, pp. 159-165, 1978.






\end{thebibliography}
\end{document}